\documentclass{article}
\usepackage[utf8]{inputenc}
\usepackage{graphicx}
\usepackage[affil-it]{authblk}
\usepackage{float}
\usepackage{caption}
\usepackage{subcaption}
\usepackage[backend=biber,style=ieee,sorting=none]{biblatex}
\usepackage[usenames, dvipsnames]{color}
\usepackage{amsmath}
\usepackage{longtable}

\title{Spatio-temporal Symmetry - Point Groups with Time Translations}
\author[1]{Haricharan Padmanabhan} 
\author[1,2]{Maggie L. Kingsland}
\author[1]{Jason M. Munro}
\author[3]{Daniel B. Litvin}
\author[1]{Venkatraman Gopalan}

\affil[1]{Department of Materials Science and Engineering, The Pennsylvania State University, University Park, PA 16801, USA; munrojm@psu.edu (J.M.M.); vgopalan@psu.edu (V.G.)}
\affil[2]{Department of Physics, University of South Florida, Tampa, FL 33620, USA; kingslandm@mail.usf.edu}
\affil[3]{Department of Physics, The Eberly College of Science, The Pennsylvania State University, Penn State Berks, PO Box 7009, Reading, PA 19610, USA; u3c@psu.edu}

\date{Today}

\begin{document}

\maketitle

\begin{abstract}
Spatial symmetries occur in combination with temporal symmetries in a wide range of physical systems in nature, including time-periodic quantum systems typically described by the Floquet formalism.  In this context, groups formed by three-dimensional point group symmetry operations in combination with time translation operations are discussed in this work. The derivation of these 'spatio-temporal' groups from conventional point groups and their irreducible representations is outlined, followed by a complete listing. The groups are presented in a template similar to space group operations, and are visualized using a modified version of conventional stereographic projections. Simple examples of physical processes that simultaneously exhibit symmetry in space and time are identified and used to illustrate the application of spatio-temporal groups.
\end{abstract}

\section{Introduction}

Spatial symmetries are ubiquitous in nature, ranging from atoms and molecules to crystals and biological systems. The mathematical groups corresponding to these symmetries, i. e. point groups and space groups, have been listed exhaustively and in great detail \cite{Aroyo2016}, and are indispensable in the study of matter. In this work, we consider the groups formed by spatial symmetries in combination with temporal symmetries.

There are different ways in which temporal symmetries occur in physical systems. Most notably, strongly driven time-periodic quantum systems are typically described by the Floquet formalism, which involves a time-periodic Hamiltonian with its corresponding time-periodic solutions. Examples include problems that consider interaction of matter with strong electromagnetic fields, such as in high-harmonic generation of light \cite{Moiseyev1997}. Separately, an idea proposed by Wilczek et al. \cite{Wilczek2012, Shapere2012} considers time-independent Hamiltonians that spontaneously break time-translational symmetry, leading to the idea of 'time crystals'. This is a topic that has experienced a flurry of activity \cite{Choi2016a, Zhang2016a, Yao2017a} and debate \cite{Bruno2013a, Watanabe2015a} in recent years. In all these examples, with the addition of such temporal symmetries to the spatial symmetries intrinsic to these systems, it is appropriate to describe them using symmetries that combine operations in space and time, i.e. spatio-temporal symmetries, rather than conventional spatial symmetry operations. Much like in other areas of science, symmetry can be a powerful tool in the study of these systems, such as in labeling Floquet states \cite{Alon2002a}, deriving selection rules for high-harmonic generation spectra \cite{Alon1998, Ceccherini2001}, identifying symmetry-protected topological Floquet phases \cite{Morimoto2017}, deriving the form of property tensors of space-time crystals, and so on. A systematic listing of spatio-temporal groups would facilitate their use in such applications.

This paper presents the derivation and listing of groups that combine spatial operations with time-translations. While spatio-temporal groups have been previously listed \cite{Janssen1969}, they have not found widespread use, perhaps because they have not been sufficiently comprehensible to the general reader, unlike the widely used conventional point group and space group listings \cite{Aroyo2016}. In this work, the listing of groups is reformulated with the intention to remedy this problem. This includes outlining a straightforward derivation using character tables of conventional point groups, representing them using a template similar to space group operations, and devising a simple way to represent these using standard crystallographic diagrams. Furthermore, some simple examples are shown to demonstrate how these groups can be applied to physical systems. 

While the spatio-temporal groups corresponding to the 32 crystallographic point groups are listed explicitly, formulas are listed to generate the spatio-temporal groups corresponding to the non-crystallographic point groups. 

\section{Derivation}

Define a point $({\bf r}|t)$ and an operation $(R|\tau)$ in four-dimensional space-time, where ${\bf r}$ is the vector of three-dimensional spatial coordinates, $t$ is the time coordinate, $R$ is a proper or improper rotation, and $\tau$ is a time translation, such that \((R|\tau)({\bf r}|t) = (R{\bf r}|t+\tau)\). The objective is to list all possible groups of such operations. 
	
Consider the group of all spatial symmetry operations in three-dimensions, ${\bf E_s}(3)$, and the group of all time translations ${\bf E_t}(1)$. The stated objective is equivalent to listing all the subgroups of the direct product ${\bf E_s}(3)\times {\bf E_t}(1)$. The \textit{isomorphism theorem} \cite{Litvin1974} can be used to do this. Consider two groups ${\bf A}$ and ${\bf B}$, and the direct product ${\bf A}\times {\bf B}$. Choose two arbitrary normal subgroups (with different subgroups indexed by $j$), ${\bf a_j}$ and ${\bf b_j}$, of ${\bf A}$ and ${\bf B}$ respectively. Performing a coset decomposition, 
	
\begin{equation}
\begin{split}
    &{\bf A} = {\bf a_j} + A_1{\bf a_j} + A_2{\bf a_j} +...+ A_n{\bf a_j}\\
    &{\bf B} = {\bf b_j} + B_1{\bf b_j} + B_2{\bf b_j} + ... + B_n{\bf b_j}.
\end{split}
\end{equation} 
	
The isomorphism theorem states that if the factor groups ${\bf A/a_j}$ and ${\bf B/b_j}$ are isomorphic to each other, \({\bf X_j} = ({\bf a_j}|{\bf b_j})\{ (1|1), (A_1|B_1), (A_2|B_2), ... \}\) is a subgroup of ${\bf A} \times {\bf B}$. 
	
The above derivation is illustrated with an example. Consider \({\bf A} =  {\bf 4_z} = \{1,4_z,2_z,4_z^{-1}\}\), where $n_{\lambda}$ represents an anti-clockwise n-fold rotation about the $\lambda$-axis, and \({\bf B}={\bf T} =\{...-1,0,1...\}\), the set of all integral time translations, i.e. translations by integral multiples of unit time. Choosing the normal subgroups \({\bf a}={\bf 2_z}=\{1,2_z\}\) and \({\bf b}={\bf 2T}=\{...-2,0,2...\}\), it is easy to verify that the factor groups ${\bf 4_z/2_z}$ and ${\bf T/2T}$ are isomorphic to each other. Using the isomorphism theorem, \({\bf X_j}=({\bf 2_z}|{\bf 2T})\{(1|0),(4_z|1)\}\) is a subgroup of the direct product ${\bf 4_z} \times {\bf T}$. Rearranging the terms, this group can be written as $(1|{\bf 2T})\{(1|0),(4_z|1), (2_z|0), (4_z^{-1}|1)\}$. Equivalent spatio-temporal groups are defined in this work as two groups that can be transformed from one to the other either by rescaling the unit of time or by a proper spatial rotation. In the case of ${\bf X_j}$, multiplying the unit of time by a factor of $2$ gives the equivalent group $(1|{\bf T})\{(1|0),(4_z|\frac{1}{2}), (2_z|0), (4_z^{-1}|\frac{1}{2})\}$. Repeating this process with the other normal subgroups ${\bf a_j}$ and ${\bf b_j}$, all the spatio-temporal subgroups ${\bf X_j}$ of ${\bf 4_z}\times {\bf T}$ may be listed. In general, this process may still result in spatio-temporal groups that can be transformed from one to the other by a proper spatial rotation, which are hence equivalent. In this work, one group is listed from each set of such equivalent groups.
	
An alternative but mathematically equivalent approach was shown by Boyle et al. \cite{Boyle2016}. Each one-dimensional irrep of a group ${\bf G}$ is associated with a unique spatio-temporal subgroup of ${\bf G}\times {\bf T}$, where ${\bf T}$ $=\{...-1,0,1...\}$, the group of integral time translations. Given the $i^{th}$ one-dimensional irrep $\chi_i$ of a group ${\bf G}$, each element $g_{ij}$ of the irrep is mapped to $\tau_i(g_{ij})$ using $g_{ij}=\exp(2\pi i \tau_i(g_{ij}))$, and the subgroup corresponding to this irrep can be listed as 

\begin{equation}
  {\bf X_i} = \{ (g_{i1}|\tau_i(g_{i1})), (g_{i2}|\tau_i(g_{i2})), (g_{i3}|\tau_i(g_{i3})), ... \}.
\end{equation}
	
For example, consider the group \({\bf 4_z}=\{1,4_z,2_z,4_z^{-1}\}\), and its second one-dimensional irrep, \(\chi_2 = \{1,-1,1,-1\}\). The irrep can be expressed as 
\begin{equation}
\begin{split}
     \chi_2 &= \{1,-1,1,-1\}\\
     &=\{exp(2\pi i (n)), exp(2\pi i (n+\frac{1}{2})),exp(2\pi i (n)),exp(2\pi i (n+\frac{1}{2}))\}\\
     &\implies \tau_2=\{n, n+\frac{1}{2}, n, n+\frac{1}{2}\}=n\{0,\frac{1}{2},0,\frac{1}{2}\}.
\end{split}
\end{equation}

The subgroup corresponding to this irrep is then $(1|{\bf T})\{(1|0),(4_z|\frac{1}{2}), (2_z|0), (4_z^{-1}|\frac{1}{2})\}$, which is the same as that obtained using the isomorphism theorem, with the normal subgroups ${\bf 2_z}$ and ${\bf 2T}$. Running through all the one-dimensional irreps of a point group in this manner is equivalent to going through all the sets of normal subgroups. 
	
As stated by Boyle et al. \cite{Boyle2016}, the above method can be used to generate a complete list of spatio-temporal point groups, which we do by going through the one-dimensional irreps of each of the 32 crystalline point groups, as well as non-crystalline point groups. Those obtained from crystallographic point groups are explicitly listed, while formulas are listed for the spatio-temporal groups obtained from non-crystallographic point groups. Note that the crystallographic spatio-temporal groups are necessarily obtained from finite spatial groups, whereas the non-crystallographic spatio-temporal groups include both finite as well as infinite spatial groups.

\section{Listing}

The groups are listed in sets according to the underlying point groups used to generate them. Each group in the set is also assigned a serial number for identification. Positions in these are separated by a period, and from left to right represent the underlying point group of the translation group, the number of the group in the set of groups listed under a specific point group, and the overall index of the group with respect to all possible time translation groups, respectively. For example, the group 11.3.29 in Table 1 refers to the twenty ninth listed spatio-temporal group, which is the third group in the series of groups constructed from the eleventh point group (which is $\frac{4}{m}$). 

The elements of the group are expressed as $(R|\tau)$, where $R$ is a proper or improper rotation, and $\tau$ is a time translation. The standard crystallographic notation for spatial symmetry as found in the International Tables for Crystallography, Volume A \cite{Aroyo2016} is used to express the proper and improper rotations. Further, non-zero time-translations are shown in blue.
	
Because of the infinite nature of the time translation groups, they are listed using the coset representatives of their decomposition with respect to the normal subgroup of all integral time translations. For example, the group
$$(1|{\bf T})\{(1|0),(4_z|\frac{1}{2}), (2_z|0), (4_z^{-1}|\frac{1}{2})\}$$ is given by listing its coset representatives with respect to ${\bf T}$, which are
$$\{(1|0),(4_z|\frac{1}{2}), (2_z|0), (4_z^{-1}|\frac{1}{2})\}.$$ 
	
Finally, a simple method is devised to help visualize these groups, by modifying conventional point group stereographic projections. Time translations are indicated in the diagram in a manner similar to how spatial translations perpendicular to the plane are indicated in space-group diagrams. Non-zero time translations are visually indicated by numbers in blue. The spatial element associated with a time translation is located within the plane by proximity, and outside the plane using the superscript. An example is shown in Fig. 1, and the supplementary information contains stereographic projections for the remaining groups as well, listed according to their serial numbers.

\begin{figure}
	\centering
	\includegraphics[width=.4\linewidth]{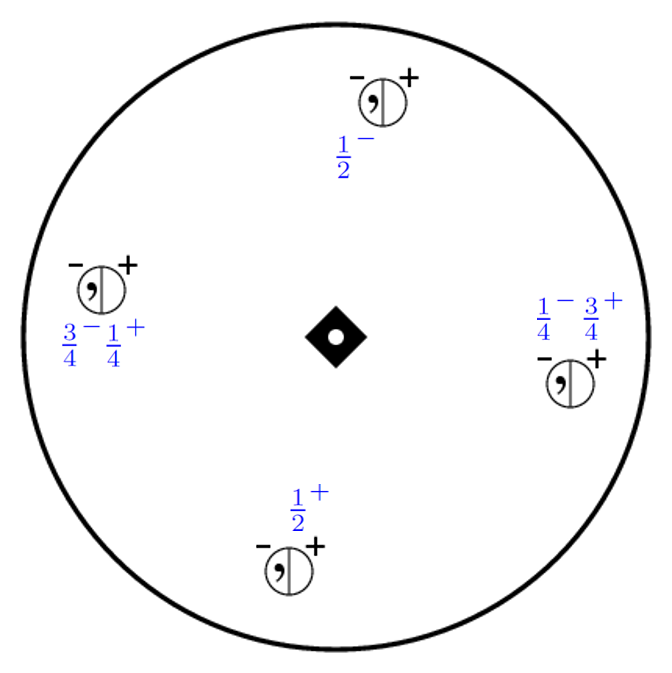}
	\captionsetup{type=figure}
	\caption{The stereographic projection of the group 11.3.29 from Table 1, which is $(1|{\bf T})\{(1|0), (4_z|\frac{1}{4}), (2_z |\frac{1}{2}), (4_z^{\, -1} | \frac{3}{4}), (\bar{1} | 0), (\bar{4}_z^{\, -1} | \frac{3}{4}), (m_z | \frac{1}{2}), (\bar{4_z} | \frac{1}{4}) $. The time translations are indicated visually on the stereographic projection with blue fractions, and the $+$ ($-$) superscript is used to specify that the time translation is associated with a spatial element above (below) the plane.}
	\label{fig1:sub2}
\end{figure}
	
In addition to the explicit listing of spatio-temporal groups obtained from the 32 crystallographic point groups, formulas to generate spatio-temporal groups corresponding to the non-crystallographic point groups have also been listed in Table 2. The method of listing and notation used is similar to that of the crystallographic spatio-temporal groups.

\section{Examples}

Spatio-temporal symmetries are seen in many complex physical systems, as outlined in the introduction, but the simplest example of one is the ubiquitous classical harmonic oscillator. Indeed, its temporal symmetry is simple enough that it is universally described using just spatial groups, as in molecular and lattice vibrations. It does however, exhibit non-trivial spatio-temporal symmetry. Consider an oscillator which is described by the equation $x=x_0\ sin(\omega t + \phi)$, where $\omega = 2\pi/\tau$, with $\tau$ being the time-period of oscillation. It is clear that applying the spatial operation $m_x$ ($x \rightarrow -x$) in combination with the time translation operation $\frac{\tau}{2}$ ($t \rightarrow t + \frac{\tau}{2}$) leaves the equation invariant. In other words, $(m_x|\frac{\tau}{2})$ is a symmetry of this system. Since it has no other non-trivial spatio-temporal symmetries, the spatio-temporal group that describes this oscillation is $(1|\boldsymbol{ \tau})\{(1|0),(m_x|\frac{\tau}{2})\}$. The equivalent spatio-temporal group $(1|{\bf T})\{(1|0),(m_x|\frac{1}{2})\}$ is obtained by dividing the unit of time by $\tau$. This corresponds to the group $4.2.7$ in Table 1. Other harmonic systems such as plane waves, molecular vibrations, etc. exhibit similar spatio-temporal symmetries, with time translations of $\frac{\tau}{2}$ coupled to spatial symmetry operations. 

\begin{figure}
	\centering
	\includegraphics[width=.4\linewidth]{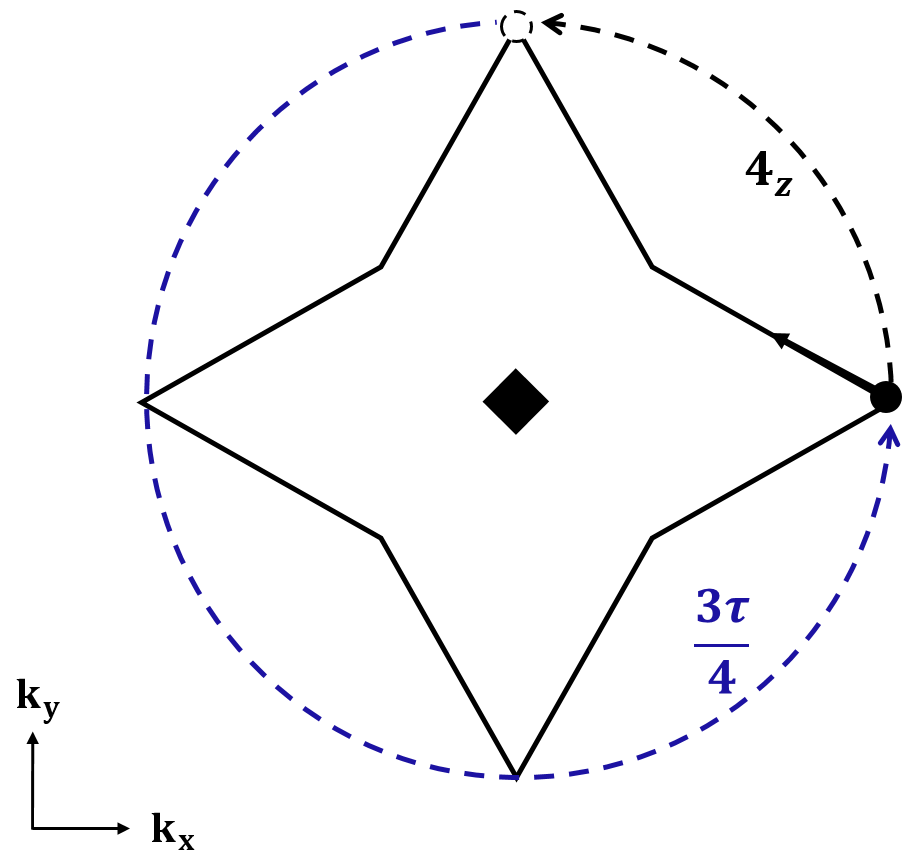}
	\captionsetup{type=figure}
   	 \caption{A schematic of the motion of an electron in k-space along the cross-section of a Fermi surface, under the influence of a magnetic field in the z-direction. One of the spatio-temporal symmetries of this motion is shown.}
	\label{fig2:sub2}
\end{figure}
	
More complex harmonic systems can exhibit higher order symmetries. A particular physical example of this is the motion in k-space, of electrons in a solid, within the semiclassical model of electron dynamics \cite{Ashcroft}. Electrons under a uniform magnetic field follow an orbit in k-space given by the intersection of the Fermi surface with planes normal to the magnetic field. Depending on the symmetry of the crystal and the direction of the magnetic field, these orbits can have different symmetries. Consider the schematic orbit shown in Fig. 2. The operation $(4_z|\frac{3\tau}{4})$  is a symmetry of this motion, as described in the figure. Using this as a generator, the spatio-temporal group $(1|{\bf T})\{(1|0), (4_z|\frac{3\tau}{4}), (2_z|\frac{\tau}{2}), (4_z^{-1}|\frac{\tau}{4})\}$ is obtained, which is equivalent to the group $9.3.23$ in Table 1. 

Much like conventional spatial symmetry, spatio-temporal symmetry can also be applied to derive properties of physical systems. For example, the selection rules for high-harmonic generation spectra can be derived using spatio-temporal symmetry. This has been shown in previous works \cite{Beswick1993a, Alon1998, Ceccherini2001}, and the simplest case of this process is outlined below. It can be shown \cite{Beswick1993a} that under the influence of linearly polarized light ${\bf E} = E_o cos(\omega t + \phi) {\bf x}$, (using the semiclassical picture of light-matter interaction) the probability to generate the $n^{th}$ harmonic from a system in a Floquet state $\psi_{\epsilon}=exp(-i\epsilon t \hbar)\phi_{\epsilon}$ is given by

\begin{equation}
    \sigma_{\epsilon}^{(n)} \propto n^4|\langle\langle\phi_{\epsilon}|\hat{\mu} e^{-in\omega t}|\phi_{\epsilon}\rangle\rangle|^2
,
\end{equation} 

where $\hat{\mu}$ is the dipole moment operator, $\omega$ is the frequency of the incident light, and $\langle \langle .. \rangle \rangle$ stands for integration over spatial variables and time. Note that the electric field, and hence the Hamiltonian is invariant under the spatio-temporal symmetry operation $(m_x|\frac{\tau}{2})$. Hence, if there is no degeneracy in the Floquet states, $|\phi_{\epsilon}\rangle\rangle$ are simultaneous eigenstates of the Floquet Hamiltonian as well as elements of the group generated by the operation $(m_x|\frac{\tau}{2})$, i. e. $(1|\bf{\tau})\{(1|0), (m_x|\frac{\tau}{2})\}$, which is equivalent to the group $4.2.7$ in the listing. Being a second order symmetry operation, $(m_x|\frac{\tau}{2})$ has eigenvalues of $\pm 1$. Applying a spatio-temporal coordinate transformation $\hat{M} = (m_x|\frac{\tau}{2})$ to the matrix element in (4),

\begin{equation}
    n^4|\langle\langle\phi_{\epsilon}|\hat{\mu} e^{-in\omega t}|\phi_{\epsilon}\rangle\rangle|^2 = n^4|\langle\langle\hat{M}\phi_{\epsilon}|\hat{M}\hat{\mu} e^{-in\omega t}\hat{M^{-1}}|\hat{M}\phi_{\epsilon}\rangle\rangle|^2 \neq 0
\end{equation}

for a non-vanishing probability of obtaining the n$^{th}$ harmonic. Using the eigenvalues of $\hat{M}$ given by $\hat{M} |\phi_{\epsilon}\rangle\rangle= \pm |\phi_{\epsilon}\rangle\rangle$, it is inferred that

\begin{equation}
    \hat{\mu}(x) e^{-in\omega t}=\hat{M}\hat{\mu}(x) e^{-in\omega t}\hat{M^{-1}}=\hat{\mu}(-x)e^{(-in\omega (t + \tau/2))}.
\end{equation}

It is clear from (6) that the matrix element is non-vanishing only for odd $n$, resulting in the selection rule that under linearly polarized light, only the harmonics given by odd $n$ are allowed in this Floquet state. Such selection rules can be derived for more complex systems, such as crystals with non-trivial spatial symmetry, and elliptically polarized incident light.

\section{Extension from time to a time-like coordinate}

A parallel can be drawn between spatio-temporal groups and the distortion antisymmetry groups formulated by VanLeeuwen and Gopalan \cite{VanLeeuwen2015}. Certain physical systems that can be described by spatio-temporal groups with time translations of $\frac{\tau}{2}$ can also be described by distortion groups obtained from the corresponding point group. The simple harmonic oscillator is a simple example of this. By parameterizing the oscillation using $\lambda$, where $-1 < \lambda < +1$, and $\lambda=0$ defines the equilibrium position, the distortion group of this system is $m^*_x = \{1,m^*_x\}$, while the spatio-temporal group is  $(1|{\bf T})\{(1|0),(m_x|\frac{\tau}{2})\}$. Furthermore, borrowing from the concept of distortion symmetry, where $\lambda$ is a 'time-like' coordinate rather than the time-coordinate itself, these point groups with time-translations can be extended to include 'time-like' translations, or 'distortion' translations.  This opens up the possibility of describing a whole range of problems using these groups, such as diffusion in materials, which may not be periodic in time, but still exhibit symmetry in a 'time-like' coordinate. 
	
Finally, in order to extend the scope of spatio-temporal symmetry, additional groups can be derived using space groups, magnetic space groups, and by considering time reversal symmetry. These groups could describe spatial translation and time reversal symmetries in addition to the point group operations and time translations described by the listing in this work, pushing the possible boundaries of application.

\pagebreak
\section*{Table 1 - List of crystallographic spatio-temporal groups}

This table includes a complete listing of the crystallographic spatio-temporal point groups. The first column assigns a serial number for each spatio-temporal group. The second column specifies which point group the corresponding spatio-temporal group was derived from. The third column lists a coset representative of each non-equivalent spatio-temporal point group, as described in the section 'Listing'. Each element $(R|\tau)$  in this column consists of a spatial component $R$ and a time translation $\tau$, with non-zero time-translations shown in blue. 

For the point groups with three-fold and six-fold axial symmetry, the following convention is used - the axis '1' is chosen to be in the in-plane horizontal direction, and the axis 'x' makes an angle of $-\frac{\pi}{6}$ with respect to it. The sets of axes 1, 2, and 3, and x, y, and xy are each generated by threefold rotations about the out-of-plane direction.

\begin{longtable}{lll}
Serial Number   & Point Group   & Spatio-temporal Group   \\
\hline
 1.1.1      & 1             &$ ( 1 | 0) $ \\
 2.1.2		& $\bar{1}$     &$ ( 1 | 0) \; \;  (\bar{1} | 0) $     \\
 2.2.3		& 		        &$ ( 1 | 0) \; \;  (\bar{1} | \textcolor{blue}{ \frac{1}{2}}) $     \\
 3.1.4		& 2 		    &$ ( 1 | 0) \; \;  (2_y | 0) $ \\
 3.2.5		&  				&$  ( 1 | 0) \; \; (2_y | \textcolor{blue}{ \frac{1}{2}}) $ \\ 
 4.1.6		&m				&$ ( 1 | 0) \; \;  (m_y | 0) $ \\
 4.2.7		&				&$ ( 1 | 0) \; \;  (m_y | \textcolor{blue}{\frac{1}{2}}) $ \\ 
 5.1.8		&2/m			&$ ( 1 | 0) \; \; (2_y | 0)  \; \; (\bar{1}  | 0)  \; \; (m_y | 0) $ \\
 5.2.9		&				&$ ( 1 | 0) \; \;  (2_y | \textcolor{blue}{\frac{1}{2}}) \; \; (\bar{1}  | 0) \; \; (m_y | \textcolor{blue}{ \frac{1}{2}}) $ \\
 5.3.10		&				&$ ( 1 | 0) \; \;  (2_y | 0) \; \; (\bar{1}  | \textcolor{blue}{ \frac{1}{2} }) \; \; (m_y | \textcolor{blue}{ \frac{1}{2}}) $ \\
 5.4.11		&				&$ ( 1 | 0) \; \;   (2_y |  \textcolor{blue}{ \frac{1}{2}}) \; \; (\bar{1}  | \textcolor{blue}{ \frac{1}{2}}) \; \; (m_y | 0) $ \\ 
 6.1.12		&222			&$ ( 1 | 0) \; \;   (2_z | 0)  \; \; (2_y | 0) \; \; (2_x | 0) $ \\ 
 6.2.13		&				&$ ( 1 | 0) \; \; (2_z | 0)  \; \; (2_y | \textcolor{blue}{ \frac{1}{2}}) \; \; (2 _x | \textcolor{blue}{ \frac{1}{2}}) $ \\
 7.1.14		&mm2			&$  ( 1 | 0) \; \;  (2_z | 0)  \; \; (m_y | 0) \; \; (m_x | 0) $ \\
 7.2.15		&				&$  ( 1 | 0) \; \;  (2_z | 0)  \; \; (m_y | \textcolor{blue}{ \frac{1}{2}}) \; \; (m_x | \textcolor{blue}{ \frac{1}{2}}) $ \\
 7.3.16		&				&$   ( 1 | 0) \; \; (2_z | \textcolor{blue}{ \frac{1}{2}}) \; \; (m_y | 0) \; \; (m_x | \textcolor{blue}{ \frac{1}{2}}) $ \\ 
 8.1.17		&mmm			&$  ( 1 | 0) \; \;  (2_z | 0) \; \; (2_y | 0)  \; \; (2_x | 0)  \; \; (\bar{1} | 0)  \; \; (m_z | 0) $ \\ 
 			&				&$ (m_y | 0) \; \; (m_x | 0) $ \\
 8.2.18		&				&$ ( 1 | 0) \; \;   (2_z | 0) \; \; (2_y | \textcolor{blue}{ \frac{1}{2}}) \; \; (2_x | \textcolor{blue}{\frac{1}{2}}) \; \; (\bar{1} | 0) \; \; (m_z | 0) $ \\
 			&				&$ (m_y | \textcolor{blue}{ \frac{1}{2}}) \; \; (m_x | \textcolor{blue}{ \frac{1}{2}}) $ \\
 8.3.19		&				& $  ( 1 | 0) \; \;  (2_z | 0) \; \; (2_y | 0) \; \; (2_x | 0) \; \; (\bar{1} | \textcolor{blue}{ \frac{1}{2}}) \; \; (m_z | \textcolor{blue}{ \frac{1}{2}}) $\\
 			&				&$ (m_y | \textcolor{blue}{ \frac{1}{2}}) \; \; (m_x | \textcolor{blue}{ \frac{1}{2}}) $ \\
 8.4.20		&				&$  ( 1 | 0) \; \;   (2_z | 0) \; \; (2_y | \textcolor{blue}{ \frac{1}{2}}) \; \; (2_x | \textcolor{blue}{ \frac{1}{2}}) \; \; (\bar{1} | \textcolor{blue}{ \frac{1}{2}}) \; \; (m_z | \textcolor{blue}{ \frac{1}{2}}) $\\
 			&				&$ (m_y | 0) \; \; (m_x | 0) $ \\
 9.1.21		&4				&$ ( 1 | 0) \; \;   (4_z | 0) \; \; (2_z | 0) \; \; (4_z^{\, -1} | 0) $ \\
 9.2.22		&				&$  ( 1 | 0) \; \;  (4_z | \textcolor{blue}{ \frac{1}{2}}) \; \; (2_z | 0) \; \; (4_z^{\, -1} | \textcolor{blue}{ \frac{1}{2}}) $ \\
 9.3.23		&				&$  ( 1 | 0) \; \;  (4_z | \textcolor{blue}{ \frac{1}{4}}) \; \; (2_z | \textcolor{blue}{ \frac{1}{2}}) \; \; (4_z^{\, -1} | \textcolor{blue}{ \frac{3}{4}}) $ \\
 10.1.24	&$\bar{4}$		&$ ( 1 | 0) \; \;   (\bar{4_z} | 0) \; \; (2_z | 0) \; \; (\bar{4}_z^{\, -1} | 0) $ \\
 10.2.25	&				&$ ( 1 | 0) \; \;   (\bar{4_z} | \textcolor{blue}{ \frac{1}{2}}) \; \; (2_z | 0) \; \; (\bar{4}_z^{\, -1} | \textcolor{blue}{ \frac{1}{2}}) $ \\
 10.3.26	&				&$ ( 1 | 0) \; \;   (\bar{4_z} | \textcolor{blue}{ \frac{1}{4}}) \; \; (2_z | \textcolor{blue}{ \frac{1}{2}}) \; \; (\bar{4}_z^{\, -1} | \textcolor{blue}{ \frac{3}{4}}) $ \\
 11.1.27	&4/m				&$ ( 1 | 0) \; \;   (4_z | 0) \; \; (2_z | 0) \; \; (4_z^{\, -1} | 0) \; \; (\bar{1} | 0) \; \; (\bar{4}_z^{\, -1} | 0) $ \\
 			&				&$ (m_z | 0) \; \; (\bar{4_z} | 0) $ \\
 11.2.28	&				&$  ( 1 | 0) \; \;  (4_z | \textcolor{blue}{ \frac{1}{2}}) \; \; (2_z | 0) \; \; (4_z^{\, -1} | \textcolor{blue}{ \frac{1}{2}}) \; \; (\bar{1} | 0) \; \; (\bar{4}_z^{\, -1} | \textcolor{blue}{ \frac{1}{2}}) $\\
 			&				&$ (m_z | 0) \; \; (\bar{4_z} | \textcolor{blue}{ \frac{1}{2}}) $ \\
 11.3.29	&				&$  ( 1 | 0) \; \;  (4_z | \textcolor{blue}{ \frac{1}{4}}) \; \; (2_z | \textcolor{blue}{ \frac{1}{2}}) \; \; (4_z^{\, -1} | \textcolor{blue}{ \frac{3}{4}}) \; \; (\bar{1} | 0) \; \; (\bar{4}_z^{\, -1} | \textcolor{blue}{ \frac{3}{4}}) $\\
 			&				&$ (m_z | \textcolor{blue}{ \frac{1}{2}}) \; \; (\bar{4_z} | \textcolor{blue}{ \frac{1}{4}}) $ \\
11.4.30		&				&$   ( 1 | 0) \; \; (4_z | 0) \; \; (2_z | 0) \; \; (4_z^{\, -1} | 0) \; \; (\bar{1} | \textcolor{blue}{ \frac{1}{2}}) \; \; (\bar{4}_z^{\, -1} | \textcolor{blue}{ \frac{1}{2}}) $\\
 			&				&$ (m_z | \textcolor{blue}{ \frac{1}{2}}) \; \; (\bar{4_z} | \textcolor{blue}{ \frac{1}{2}}) $ \\
 11.5.31	&				&$  ( 1 | 0) \; \;  (4_z | \textcolor{blue}{ \frac{1}{2}}) \; \; (2_z | 0) \; \; (4_z^{\, -1} | \textcolor{blue}{ \frac{1}{2}}) \; \; (\bar{1} | \textcolor{blue}{ \frac{1}{2}}) \; \; (\bar{4}_z^{\, -1} | 0) $ \\
 			&				&$ (m_z | \textcolor{blue}{ \frac{1}{2}}) \; \; (\bar{4_z} | 0) $ \\
11.6.32		&				&$  ( 1 | 0) \; \;  (4_z | \textcolor{blue}{ \frac{1}{4}}) \; \; (2_z | \textcolor{blue}{ \frac{1}{2}}) \; \; (4_z^{\, -1} | \textcolor{blue}{ \frac{3}{4}}) \; \; (\bar{1} | \textcolor{blue}{ \frac{1}{2}}) \; \; (\bar{4}_z^{\, -1} | \textcolor{blue}{ \frac{1}{4}}) $ \\
 			&				&$ (m_z | 0) \; \; (\bar{4_z} | \textcolor{blue}{ \frac{3}{4}}) $ \\
 12.1.33		&422				& $  ( 1 | 0) \; \; (4_z | 0) \; \; (4_z^{\, -1} | 0) \; \; (2_z | 0) \; \; (2_y | 0) \; \; (2_x | 0) $ \\
 			&				&$ (2_{xy} | 0) \; \; (2_{-xy} | 0) $ \\
 12.2.34	&				&$  ( 1 | 0) \; \; (4_z | 0) \; \; (4_z^{\, -1} | 0) \; \; (2_z | 0) \; \; (2_y | \textcolor{blue}{ \frac{1}{2}}) \; \; (2_x | \textcolor{blue}{ \frac{1}{2}}) $ \\
 			&				&$ (2_{xy} | \textcolor{blue}{ \frac{1}{2}}) \; \; (2_{-xy} | \textcolor{blue}{ \frac{1}{2}}) $ \\
 12.3.35	&				&$ ( 1 | 0) \; \;  (4_z | \textcolor{blue}{ \frac{1}{2}}) \; \; (4_z^{\, -1} | \textcolor{blue}{ \frac{1}{2}}) \; \; (2_z | 0) \; \; (2_y | \textcolor{blue}{ \frac{1}{2}}) \; \; (2_x | \textcolor{blue}{ \frac{1}{2}}) $ \\
 			&				&$ (2_{xy} | 0) \; \; (2_{-xy} | 0) $ \\ 
 13.1.36		&4mm			&$ ( 1 | 0) \; \;   (4_z | 0) \; \; (4_z^{\, -1} | 0) \; \; (2_z | 0) \; \; (m_x | 0) \; \; (m_y | 0) $ \\
 			&				&$ (m_{xy} | 0) \; \; (m_{-xy} | 0) $ \\
 13.2.37		&				&$ ( 1 | 0) \; \;  (4_z | 0) \; \; (4_z^{\, -1} | 0) \; \; (2_z | 0) \; \; (m_x | \textcolor{blue}{ \frac{1}{2}}) \; \; (m_y | \textcolor{blue}{ \frac{1}{2}}) $ \\
 			&				&$ (m_{xy} | \textcolor{blue}{ \frac{1}{2}}) \; \; (m_{-xy} | \textcolor{blue}{ \frac{1}{2}}) $ \\
 13.3.38	&				&$ ( 1 | 0) \; \;  (4_z | \textcolor{blue}{ \frac{1}{2}}) \; \; (4_z^{\, -1} | \textcolor{blue}{ \frac{1}{2}}) \; \; (2_z | 0) \; \; (m_x | \textcolor{blue}{ \frac{1}{2}}) \; \; (m_y | \textcolor{blue}{ \frac{1}{2}}) $ \\
 			&				&$ (m_{xy} | 0) \; \; (m_{-xy} | 0) $ \\ 
 14.1.39		&$\bar{4}$2m		&$  ( 1 | 0) \; \; (\bar{4_z} | 0) \; \; (\bar{4}_z^{\, -1} | 0) \; \; (2_z | 0) \; \; (2_y | 0) \; \; (2_x | 0) $ \\
 			&				&$ (m_{xy} | 0) \; \; (m_{-xy} | 0) $ \\
 14.2.40	&				&$  ( 1 | 0) \; \; (\bar{4_z} | 0) \; \; (\bar{4}_z^{\, -1} | 0) \; \; (2_z | 0) \; \; (2_y | \textcolor{blue}{ \frac{1}{2}}) \; \; (2_x | \textcolor{blue}{ \frac{1}{2}}) $ \\
 			&				&$ (m_{xy} | \textcolor{blue}{ \frac{1}{2}}) \; \; (m_{-xy} | \textcolor{blue}{ \frac{1}{2}}) $ \\
 14.3.41	&				&$  ( 1 | 0) \; \; (\bar{4_z} | \textcolor{blue}{ \frac{1}{2}}) \; \; (\bar{4}_z^{\, -1} | \textcolor{blue}{ \frac{1}{2}}) \; \; (2_z | 0)  \; \; (2_y | 0) \; \; (2_x | 0) $ \\
 			&				&$ (m_{xy} | \textcolor{blue}{ \frac{1}{2}}) \; \; (m_{-xy} | \textcolor{blue}{ \frac{1}{2}}) $ \\
 14.4.42	&				&$ ( 1 | 0) \; \;  (\bar{4_z} | \textcolor{blue}{ \frac{1}{2}}) \; \; (\bar{4}_z^{\, -1} | \textcolor{blue}{ \frac{1}{2}}) \; \; (2_z | 0) \; \; (2_y | \textcolor{blue}{ \frac{1}{2}}) \; \; (2_x | \textcolor{blue}{ \frac{1}{2}}) $ \\
 			&				&$ (m_{xy} | 0) \; \; (m_{-xy} | 0) $ \\ 
 15.1.43	&4/mmm			&$  ( 1 | 0) \; \;  (4_z | 0) \; \; (4_z^{\, -1} | 0) \; \; (2_z | 0) \; \; (2_y | 0) \; \; (2_x | 0) $ \\
 			&				&$ (2_{xy} | 0) \; \; (2_{-xy} | 0) \; \; (\bar{1} | 0) \; \; (\bar{4_z} | 0) \; \; (\bar{4}_z^{\, -1} | 0) \; \; (m_z | 0) $ \\
			&				&$ (m_y | 0) \; \; (m_x | 0) \; \; (m_{xy} | 0) \; \; (m_{-xy} | 0) $ \\
 15.2.44	&				&$  ( 1 | 0) \; \;  (4_z | 0) \; \; (4_z^{\, -1} | 0) \; \; (2_z | 0) \; \; (2_y | \textcolor{blue}{ \frac{1}{2}}) \; \; (2_x | \textcolor{blue}{ \frac{1}{2}}) $ \\
 			&				&$ (2_{xy} | \textcolor{blue}{ \frac{1}{2}}) \; \; (2_{-xy} | \textcolor{blue}{ \frac{1}{2}}) \; \; (\bar{1} | 0) \; \; (\bar{4_z} | 0) \; \; (\bar{4}_z^{\, -1} | 0) \; \; (m_z | 0) $ \\
			&				&$ (m_y | \textcolor{blue}{ \frac{1}{2}}) \; \; (m_x | \textcolor{blue}{ \frac{1}{2}}) \; \; (m_{xy} | \textcolor{blue}{ \frac{1}{2}}) \; \; (m_{-xy} | \textcolor{blue}{ \frac{1}{2}}) $ \\
 15.3.45	&				&$  ( 1 | 0) \; \;  (4_z | \textcolor{blue}{ \frac{1}{2}}) \; \; (4_z^{\, -1} | \textcolor{blue}{ \frac{1}{2}}) \; \; (2_z | 0) \; \; (2_y | \textcolor{blue}{ \frac{1}{2}}) \; \; (2_x | \textcolor{blue}{ \frac{1}{2}}) $ \\
 			&				&$ (2_{xy} | 0) \; \; (2_{-xy} | 0) \; \; (\bar{1} | 0) \; \; (\bar{4_z} | \textcolor{blue}{ \frac{1}{2}}) \; \; (\bar{4}_z^{\, -1} | \textcolor{blue}{ \frac{1}{2}}) \; \; (m_z | 0) $ \\
			&				&$ (m_y | \textcolor{blue}{ \frac{1}{2}}) \; \; (m_x | \textcolor{blue}{ \frac{1}{2}}) \; \; (m_{xy} | 0) \; \; (m_{-xy} | 0) $ \\
 15.4.46	&				&$  ( 1 | 0) \; \;  (4_z | 0) \; \; (4_z^{\, -1} | 0) \; \; (2_z | 0) \; \; (2_y | 0) \; \; (2_x | 0) $ \\
 			&				&$ (2_{xy} | 0) \; \; (2_{-xy} | 0) \; \; (\bar{1} | \textcolor{blue}{ \frac{1}{2}}) \; \; (\bar{4_z} | \textcolor{blue}{ \frac{1}{2}}) \; \; (\bar{4}_z^{\, -1} | \textcolor{blue}{ \frac{1}{2}}) \; \; (m_z | \textcolor{blue}{ \frac{1}{2}}) $ \\
			&				&$ (m_y | \textcolor{blue}{ \frac{1}{2}}) \; \; (m_x | \textcolor{blue}{ \frac{1}{2}}) \; \; (m_{xy} | \textcolor{blue}{ \frac{1}{2}}) \; \; (m_{-xy} | \textcolor{blue}{ \frac{1}{2}}) $ \\
 15.5.47	&				&$  ( 1 | 0) \; \;  (4_z | 0) \; \; (4_z^{\, -1} | 0) \; \; (2_z | 0) \; \; (2_y | \textcolor{blue}{ \frac{1}{2}}) \; \; (2_x | \textcolor{blue}{ \frac{1}{2}}) $ \\
 			&				&$ (2_{xy} | \textcolor{blue}{ \frac{1}{2}}) \; \; (2_{-xy} | \textcolor{blue}{ \frac{1}{2}}) \; \; (\bar{1} | \textcolor{blue}{ \frac{1}{2}}) \; \; (\bar{4_z} | \textcolor{blue}{ \frac{1}{2}}) \; \; (\bar{4}_z^{\, -1} | \textcolor{blue}{ \frac{1}{2}}) \; \; (m_z | \textcolor{blue}{ \frac{1}{2}}) $ \\
			&				&$ (m_y | 0) \; \; (m_x | 0) \; \; (m_{xy} | 0) \; \; (m_{-xy} | 0) $ \\
 15.6.48	&				&$ ( 1 | 0) \; \;   (4_z | \textcolor{blue}{ \frac{1}{2}}) \; \; (4_z^{\, -1} | \textcolor{blue}{ \frac{1}{2}}) \; \; (2_z | 0) \; \; (2_y | 0) \; \; (2_x | 0) $ \\
 			&				&$ (2_{xy} | \textcolor{blue}{ \frac{1}{2}}) \; \; (2_{-xy} | \textcolor{blue}{ \frac{1}{2}}) \; \; (\bar{1} | \textcolor{blue}{ \frac{1}{2}}) \; \; (\bar{4_z} | 0) \; \; (\bar{4}_z^{\, -1} | 0) \; \; (m_z | \textcolor{blue}{ \frac{1}{2}}) $ \\
			&				&$ (m_y | \textcolor{blue}{ \frac{1}{2}}) \; \; (m_x | \textcolor{blue}{ \frac{1}{2}}) \; \; (m_{xy} | 0) \; \; (m_{-xy} | 0) $ \\
 16.1.49	&3				&$ ( 1 | 0) \; \;   (3_z | 0) \; \; (3_z^{\, -1} | 0) $ \\
 16.2.50	&				&$  ( 1 | 0) \; \; (3_z | \textcolor{blue}{ \frac{1}{3}}) \; \; (3_z^{\, -1} | \textcolor{blue}{ \frac{2}{3}}) $ \\
 17.1.51	&$\bar{3}$		&$  ( 1 | 0) \; \; (3_z | 0) \; \; (3_z^{\, -1} | 0) \; \; (\bar{1} | 0) \; \; (\bar{3}_z^{\, -1} | 0) \; \; (\bar{3}_z | 0) $ \\
 17.2.52	&				&$  ( 1 | 0) \; \; (3_z | \textcolor{blue}{ \frac{1}{3}}) \; \; (3_z^{\, -1} | \textcolor{blue}{ \frac{2}{3}}) \; \; (\bar{1} | 0) \; \; (\bar{3}_z^{\, -1} | \textcolor{blue}{ \frac{2}{3}}) \; \; (\bar{3}_z | \textcolor{blue}{ \frac{1}{3}}) $ \\
 17.3.53	&				&$ ( 1 | 0) \; \;  (3_z | 0) \; \; (3_z^{\, -1} | 0) \; \; (\bar{1} | \textcolor{blue}{ \frac{1}{2}}) \; \; (\bar{3}_z^{\, -1} | \textcolor{blue}{ \frac{1}{2}}) \; \; (\bar{3}_z | \textcolor{blue}{ \frac{1}{2}}) $ \\
 17.4.54	&				&$  ( 1 | 0) \; \; (3_z | \textcolor{blue}{ \frac{1}{3}}) \; \; (3_z^{\, -1} | \textcolor{blue}{ \frac{2}{3}}) \; \; (\bar{1} | \textcolor{blue}{ \frac{1}{2}}) \; \; (\bar{3}_z^{\, -1} | \textcolor{blue}{ \frac{1}{6}}) \; \; (\bar{3}_z | \textcolor{blue}{ \frac{5}{6}}) $ \\
 18.1.55	&32				&$ ( 1 | 0) \; \;  (3_z | 0) \; \; (3_z^{\, -1} | 0) \; \; (2_x | 0) \; \; (2_y | 0) \; \; (2_{xy} | 0) $ \\
 18.2.56	&				&$ ( 1 | 0) \; \;  (3_z | 0) \; \; (3_z^{\, -1} | 0) \; \; (2_x | \textcolor{blue}{ \frac{1}{2}}) \; \; (2_y | \textcolor{blue}{ \frac{1}{2}}) \; \; (2_{xy} | \textcolor{blue}{ \frac{1}{2}}) $ \\ 
 19.1.57	&3m				&$ ( 1 | 0) \; \;  (3_z | 0) \; \; (3_z^{\, -1} | 0) \; \; (m_x | 0) \; \; (m_y | 0) \; \; (m_{xy} | 0) $ \\
 19.2.58	&				&$  ( 1 | 0) \; \; (3_z | 0) \; \; (3_z^{\, -1} | 0) \; \; (m_x | \textcolor{blue}{ \frac{1}{2}}) \; \; (m_y | \textcolor{blue}{\frac{1}{2}}) \; \; (m_{xy} | \textcolor{blue}{ \frac{1}{2}}) $ \\ 
 20.1.59	&$\bar{3}$m		&$  ( 1 | 0) \; \;  (3_z | 0) \; \; (3_z^{\, -1} | 0) \; \; (2_1 | 0) \; \; (2_2 | 0) \; \; (2_3 | 0) $ \\
 			&				&$ (\bar{1} | 0) \; \; (\bar{3}_z | 0) \; \; (\bar{3}_z^{\, -1} | 0) \; \; (m_x | 0) \; \; (m_y | 0) \; \; (m_{xy} | 0) $ \\
 20.2.60	&				&$ ( 1 | 0) \; \;   (3_z | 0) \; \; (3_z^{\, -1} | 0) \; \; (2_1 | \textcolor{blue}{ \frac{1}{2}}) \; \; (2_2 | \textcolor{blue}{ \frac{1}{2}}) \; \; (2_3 | \textcolor{blue}{ \frac{1}{2}}) $ \\
 			&				&$ (\bar{1} | 0) \; \; (\bar{3}_z | 0) \; \; (\bar{3}_z^{\, -1} | 0) \; \; (m_x | \textcolor{blue}{ \frac{1}{2}}) \; \; (m_y | \textcolor{blue}{ \frac{1}{2}}) \; \; (m_{xy} | \textcolor{blue}{ \frac{1}{2}}) $ \\
 20.3.61		&				&$  ( 1 | 0) \; \;  (3_z | 0) \; \; (3_z^{\, -1} | 0) \; \; (2_1 | 0) \; \; (2_2 | 0) \; \; (2_3 | 0) $ \\
 			&				&$ (\bar{1} | \textcolor{blue}{ \frac{1}{2}}) \; \; (\bar{3}_z | \textcolor{blue}{ \frac{1}{2}}) \; \; (\bar{3}_z^{\, -1} | \textcolor{blue}{ \frac{1}{2}}) \; \; (m_x | \textcolor{blue}{ \frac{1}{2}}) \; \; (m_y | \textcolor{blue}{ \frac{1}{2}}) \; \; (m_{xy} | \textcolor{blue}{ \frac{1}{2}}) $ \\
 20.4.62	&				&$  ( 1 | 0) \; \;  (3_z | 0) \; \; (3_z^{\, -1} | 0) \; \; (2_1 | \textcolor{blue}{ \frac{1}{2}}) \; \; (2_2 | \textcolor{blue}{ \frac{1}{2}}) \; \; (2_3 | \textcolor{blue}{ \frac{1}{2}}) $ \\
 			&				&$ (\bar{1} | \textcolor{blue}{ \frac{1}{2}}) \; \; (\bar{3}_z | \textcolor{blue}{ \frac{1}{2}}) \; \; (\bar{3}_z^{\, -1} | \textcolor{blue}{ \frac{1}{2}}) \; \; (m_x | 0) \; \; (m_y | 0) \; \; (m_{xy} | 0) $ \\
 21.1.63	&6				&$ ( 1 | 0) \; \;  (6_z | 0) \; \; (3_z | 0) \; \; (2_z | 0) \; \; (3_z^{\, -1} | 0) \; \; (6_z^{\, -1} | 0) $ \\
 21.2.64	&				&$ ( 1 | 0) \; \;  (6_z | \textcolor{blue}{ \frac{1}{2}}) \; \; (3_z | 0) \; \; (2_z | \textcolor{blue}{ \frac{1}{2}}) \; \; (3_z^{\, -1} | 0) \; \; (6_z^{\, -1} | \textcolor{blue}{ \frac{1}{2}}) $ \\
 21.3.65	&				&$ ( 1 | 0) \; \;  (6_z | \textcolor{blue}{ \frac{1}{6}}) \; \; (3_z | \textcolor{blue}{ \frac{1}{3}}) \; \; (2_z | \textcolor{blue}{ \frac{1}{2}}) \; \; (3_z^{\, -1} | \textcolor{blue}{ \frac{2}{3}}) \; \; (6_z^{\, -1} | \textcolor{blue}{ \frac{5}{6}}) $ \\
 21.4.66	&				&$  ( 1 | 0) \; \; (6_z | \textcolor{blue}{ \frac{1}{3}}) \; \; (3_z | \textcolor{blue}{ \frac{2}{3}}) \; \; (2_z | 0) \; \; (3_z^{\, -1} | \textcolor{blue}{ \frac{1}{3}}) \; \; (6_z^{\, -1} | \textcolor{blue}{ \frac{2}{3}}) $ \\
 22.1.67	&$\bar{6}$		&$ ( 1 | 0) \; \;  (3_z | 0) \; \; (3_z^{\, -1} | 0) \; \; (m_z | 0) \; \; (\bar{6}_z | 0) \; \; (\bar{6}_z^{\, -1} | 0) $ \\
 22.2.68	&				&$ ( 1 | 0) \; \;  (3_z | \textcolor{blue}{ \frac{1}{3}}) \; \; (3_z^{\, -1} | \textcolor{blue}{ \frac{2}{3}}) \; \; (m_z | 0) \; \; (\bar{6}_z | \textcolor{blue}{ \frac{2}{3}}) \; \; (\bar{6}_z^{\, -1} | \textcolor{blue}{ \frac{1}{3}}) $ \\
 22.3.69	&				&$  ( 1 | 0) \; \; (3_z | 0) \; \; (3_z^{\, -1} | 0) \; \; (m_z | \textcolor{blue}{ \frac{1}{2}}) \; \; (\bar{6}_z | \textcolor{blue}{ \frac{1}{2}}) \; \; (\bar{6}_z^{\, -1} | \textcolor{blue}{ \frac{1}{2}}) $ \\
 22.4.70	&				&$ ( 1 | 0) \; \;  (3_z | \textcolor{blue}{ \frac{1}{3}}) \; \; (3_z^{\, -1} | \textcolor{blue}{ \frac{2}{3}}) \; \; (m_z | \textcolor{blue}{ \frac{1}{2}}) \; \; (\bar{6}_z | \textcolor{blue}{ \frac{1}{6}}) \; \; (\bar{6}_z^{\, -1} | \textcolor{blue}{ \frac{5}{6}}) $ \\
 23.1.71	&6/m				&$ ( 1 | 0) \; \;   (6_z | 0) \; \; (3_z | 0) \; \; (2_z | 0) \; \; (3_z^{\, -1} | 0) \; \; (6_z^{\, -1} | 0) $ \\
 			&				&$ (\bar{1} | 0) \; \; (\bar{3}_z^{\, -1} | 0) \; \; (\bar{6}_z^{\, -1} | 0) \; \; (m_z | 0) \; \; (\bar{6}_z | 0) \; \; (\bar{3}_z | 0) $ \\
 23.2.72		&				&$ ( 1 | 0) \; \;   (6_z | \textcolor{blue}{ \frac{1}{2}}) \; \; (3_z | 0) \; \; (2_z | \textcolor{blue}{ \frac{1}{2}}) \; \; (3_z^{\, -1} | 0) \; \; (6_z^{\, -1} | \textcolor{blue}{ \frac{1}{2}}) $ \\
 			&				&$ (\bar{1} | 0) \; \; (\bar{3}_z^{\, -1} | 0) \; \; (\bar{6}_z^{\, -1} | \textcolor{blue}{ \frac{1}{2}}) \; \; (m_z | \textcolor{blue}{ \frac{1}{2}}) \; \; (\bar{6}_z | \textcolor{blue}{ \frac{1}{2}}) \; \; (\bar{3}_z | 0) $ \\
 23.3.73	&				&$ ( 1 | 0) \; \;   (6_z | \textcolor{blue}{ \frac{1}{6}}) \; \; (3_z | \textcolor{blue}{ \frac{1}{3}}) \; \; (2_z | \textcolor{blue}{ \frac{1}{2}}) \; \; (3_z^{\, -1} | \textcolor{blue}{ \frac{2}{3}}) \; \; (6_z^{\, -1} | \textcolor{blue}{ \frac{5}{6}}) $ \\
 			&				&$ (\bar{1} | 0) \; \; (\bar{3}_z^{\, -1} | \textcolor{blue}{ \frac{2}{3}}) \; \; (\bar{6}_z^{\, -1} | \textcolor{blue}{ \frac{5}{6}}) \; \; (m_z | \textcolor{blue}{ \frac{1}{2}}) \; \; (\bar{6}_z | \textcolor{blue}{ \frac{1}{6}}) \; \; (\bar{3}_z | \textcolor{blue}{ \frac{1}{3}}) $ \\
 23.4.74	&				&$  ( 1 | 0) \; \;  (6_z | \textcolor{blue}{ \frac{1}{3}}) \; \; (3_z | \textcolor{blue}{ \frac{2}{3}}) \; \; (2_z | 0) \; \; (3_z^{\, -1} | \textcolor{blue}{ \frac{1}{3}}) \; \; (6_z^{\, -1} | \textcolor{blue}{ \frac{2}{3}}) $ \\
 			&				&$ (\bar{1} | 0) \; \; (\bar{3}_z^{\, -1} | \textcolor{blue}{ \frac{1}{3}}) \; \; (\bar{6}_z^{\, -1} | \textcolor{blue}{ \frac{2}{3}}) \; \; (m_z | 0) \; \; (\bar{6}_z | \textcolor{blue}{ \frac{1}{3}}) \; \; (\bar{3}_z | \textcolor{blue}{ \frac{2}{3}}) $ \\
 23.5.75	&				&$  ( 1 | 0) \; \;  (6_z | 0) \; \; (3_z | 0) \; \; (2_z | 0) \; \; (3_z^{\, -1} | 0) \; \; (6_z^{\, -1} | 0) $ \\
 			&				&$ (\bar{1} | \textcolor{blue}{ \frac{1}{2}}) \; \; (\bar{3}_z^{\, -1} | \textcolor{blue}{ \frac{1}{2}}) \; \; (\bar{6}_z^{\, -1} | \textcolor{blue}{ \frac{1}{2}}) \; \; (m_z | \textcolor{blue}{ \frac{1}{2}}) \; \; (\bar{6}_z | \textcolor{blue}{ \frac{1}{2}}) \; \; (\bar{3}_z | \textcolor{blue}{ \frac{1}{2}}) $ \\
 23.6.76	&				&$  (1 | 0)  \; \; (6_z | \textcolor{blue}{ \frac{1}{2}}) \; \; (3_z | 0) \; \; (2_z | \textcolor{blue}{ \frac{1}{2}}) \; \; (3_z^{\, -1} | 0) \; \; (6_z^{\, -1} | \textcolor{blue}{ \frac{1}{2}}) $ \\
 			&				&$ (\bar{1} | \textcolor{blue}{ \frac{1}{2}}) \; \; (\bar{3}_z^{\, -1} | \textcolor{blue}{ \frac{1}{2}}) \; \; (\bar{6}_z^{\, -1} | 0) \; \; (m_z | 0) \; \; (\bar{6}_z | 0) \; \; (\bar{3}_z | \textcolor{blue}{ \frac{1}{2}}) $ \\
 23.7.77	&				&$ ( 1 | 0) \; \;   (6_z | \textcolor{blue}{ \frac{1}{6}}) \; \; (3_z | \textcolor{blue}{ \frac{1}{3}}) \; \; (2_z | \textcolor{blue}{ \frac{1}{2}}) \; \; (3_z^{\, -1} | \textcolor{blue}{ \frac{2}{3}}) \; \; (6_z^{\, -1} | \textcolor{blue}{ \frac{5}{6}}) $ \\
 			&				&$ (\bar{1} | \textcolor{blue}{ \frac{1}{2}}) \; \; (\bar{3}_z^{\, -1} | \textcolor{blue}{ \frac{1}{6}}) \; \; (\bar{6}_z^{\, -1} | \textcolor{blue}{ \frac{1}{3}}) \; \; (m_z | 0) \; \; (\bar{6}_z | \textcolor{blue}{ \frac{2}{3}}) \; \; (\bar{3}_z | \textcolor{blue}{ \frac{5}{6}}) $ \\
 23.8.78	&				&$ ( 1 | 0) \; \;   (6_z | \textcolor{blue}{ \frac{1}{3}}) \; \; (3_z | \textcolor{blue}{ \frac{2}{3}}) \; \; (2_z | 0) \; \; (3_z^{\, -1} | \textcolor{blue}{ \frac{1}{3}}) \; \; (6_z^{\, -1} | \textcolor{blue}{ \frac{2}{3}}) $ \\
 			&				&$ (\bar{1} | \textcolor{blue}{ \frac{1}{2}}) \; \; (\bar{3}_z^{\, -1} | \textcolor{blue}{\frac{5}{6}}) \; \; (\bar{6}_z^{\, -1} | \textcolor{blue}{ \frac{1}{6}}) \; \; (m_z | \textcolor{blue}{ \frac{1}{2}}) \; \; (\bar{6}_z | \textcolor{blue}{ \frac{5}{6}}) \; \; (\bar{3}_z | \textcolor{blue}{ \frac{1}{6}}) $ \\
 24.1.79	&622				&$  ( 1 | 0) \; \;  (6_z | 0) \; \; (6_z^{\, -1} | 0) \; \; (3_z | 0) \; \; (3_z^{\, -1} | 0) \; \; (2_z | 0) $ \\
 			&				&$ (2_x | 0) \; \; (2_1 | 0) \; \; (2_{xy} | 0) \; \; (2_2 | 0) \; \; (2_y | 0) \; \; (2_3 | 0) $ \\
 24.2.80	&				&$  ( 1 | 0) \; \;  (6_z | 0) \; \; (6_z^{\, -1} | 0) \; \; (3_z | 0) \; \; (3_z^{\, -1} | 0) \; \; (2_z | 0) $ \\
 			&				&$ (2_x | \textcolor{blue}{ \frac{1}{2}}) \; \; (2_1 | \textcolor{blue}{ \frac{1}{2}}) \; \; (2_{xy} | \textcolor{blue}{ \frac{1}{2}}) \; \; (2_2 | \textcolor{blue}{ \frac{1}{2}}) \; \; (2_y | \textcolor{blue}{ \frac{1}{2}}) \; \; (2_3 | \textcolor{blue}{ \frac{1}{2}}) $ \\
 24.3.81	&				&$ ( 1 | 0) \; \;   (6_z | \textcolor{blue}{ \frac{1}{2}}) \; \; (6_z^{\, -1} | \textcolor{blue}{ \frac{1}{2}}) \; \; (3_z | 0) \; \; (3_z^{\, -1} | 0) \; \; (2_z | \textcolor{blue}{ \frac{1}{2}}) $ \\
 			&				&$ (2_x | 0) \; \; (2_1 | \textcolor{blue}{ \frac{1}{2}}) \; \; (2_{xy} | 0) \; \; (2_2 |\textcolor{blue}{ \frac{1}{2}}) \; \; (2_y | 0) \; \; (2_3 |\textcolor{blue}{ \frac{1}{2}}) $ \\
 25.1.82	&6mm			&$  ( 1 | 0) \; \;  (6_z | 0) \; \; (6_z^{\, -1} | 0) \; \; (3_z | 0) \; \; (3_z^{\, -1} | 0) \; \; (2_z | 0) $ \\
 			&				&$ (m_x | 0) \; \; (m_1 | 0) \; \; (m_{xy} | 0) \; \; (m_2 | 0) \; \; (m_y | 0) \; \; (m_3 | 0) $ \\
 25.2.83	&				&$  ( 1 | 0) \; \;  (6_z | 0) \; \; (6_z^{\, -1} | 0) \; \; (3_z | 0) \; \; (3_z^{\, -1} | 0) \; \; (2_z | 0) $ \\
 			&				&$ (m_x | \textcolor{blue}{ \frac{1}{2}}) \; \; (m_1 | \textcolor{blue}{ \frac{1}{2}}) \; \; (m_{xy} | \textcolor{blue}{ \frac{1}{2}}) \; \; (m_2 | \textcolor{blue}{ \frac{1}{2}}) \; \; (m_y | \textcolor{blue}{ \frac{1}{2}}) \; \; (m_3 | \textcolor{blue}{ \frac{1}{2}}) $ \\
 25.3.84	&				&$  ( 1 | 0) \; \;  (6_z | \textcolor{blue}{ \frac{1}{2}}) \; \; (6_z^{\, -1} | \textcolor{blue}{ \frac{1}{2}}) \; \; (3_z | 0) \; \; (3_z^{\, -1} | 0) \; \; (2_z | \textcolor{blue}{ \frac{1}{2}}) $ \\
 			&				&$ (m_x |0) \; \; (m_1 | \textcolor{blue}{ \frac{1}{2}}) \; \; (m_{xy} |0) \; \; (m_2 |\textcolor{blue}{ \frac{1}{2}}) \; \; (m_y | 0) \; \; (m_3 | \textcolor{blue}{ \frac{1}{2}}) $ \\ 
26.1.85		&$\bar{6}$2m		&$ ( 1 | 0) \; \;   (3_z | 0) \; \; ({3_z}^{\, -1} | 0) \; \; (2_x | 0) \; \; (2_{xy} | 0) \; \; (2_y | 0) $ \\
 			&				&$ (m_z | 0) \; \; (\bar{6}_z | 0) \; \; (\bar{6}_z^{\, -1} | 0) \; \; (m_1 | 0) \; \; (m_2 | 0) \; \; (m_3 | 0) $ \\
 26.2.86	&				&$ ( 1 | 0) \; \;   (3_z | 0) \; \; ({3_z}^{\, -1} | 0) \; \; (2_x | \textcolor{blue}{ \frac{1}{2}}) \; \; (2_{xy} | \textcolor{blue}{ \frac{1}{2}}) \; \; (2_y | \textcolor{blue}{ \frac{1}{2}}) $ \\
 			&				&$ (m_z | 0) \; \; (\bar{6}_z | 0) \; \; (\bar{6}_z^{\, -1} | 0) \; \; (m_1 | \textcolor{blue}{ \frac{1}{2}}) \; \; (m_2 | \textcolor{blue}{ \frac{1}{2}}) \; \; (m_3 | \textcolor{blue}{ \frac{1}{2}}) $ \\
 26.3.87	&				&$  ( 1 | 0) \; \;  (3_z | 0) \; \; ({3_z}^{\, -1} | 0) \; \; (2_x | 0) \; \; (2_{xy} | 0) \; \; (2_y | 0) $ \\
 			&				&$ (m_z | \textcolor{blue}{ \frac{1}{2}}) \; \; (\bar{6}_z | \textcolor{blue}{ \frac{1}{2}}) \; \; (\bar{6}_z^{\, -1} | \textcolor{blue}{ \frac{1}{2}}) \; \; (m_1 | \textcolor{blue}{ \frac{1}{2}}) \; \; (m_2 | \textcolor{blue}{ \frac{1}{2}}) \; \; (m_3 | \textcolor{blue}{ \frac{1}{2}}) $ \\
 26.4.88	&				&$  ( 1 | 0) \; \;  (3_z | 0) \; \; ({3_z}^{\, -1} | 0) \; \; (2_x | \textcolor{blue}{ \frac{1}{2}}) \; \; (2_{xy} | \textcolor{blue}{ \frac{1}{2}}) \; \; (2_y | \textcolor{blue}{ \frac{1}{2}}) $ \\
 			&				&$ (m_z | \textcolor{blue}{ \frac{1}{2}}) \; \; (\bar{6}_z | \textcolor{blue}{ \frac{1}{2}}) \; \; (\bar{6}_z^{\, -1} | \textcolor{blue}{ \frac{1}{2}}) \; \; (m_1 | 0) \; \; (m_2 | 0) \; \; (m_3 | 0) $ \\ 
 27.1.89		&6/mmm			&$   ( 1 | 0) \; \; (6_z | 0) \; \; (6_z^{\, -1} | 0) \; \; (3_z | 0) \; \; (3_z^{\, -1} | 0) \; \; (2_z | 0) $ \\
 			&				&$ (2_x | 0) \; \; (2_1 | 0) \; \; (2_{xy} | 0) \; \; (2_2 | 0) \; \; (2_y | 0) \; \; (2_3 | 0) $ \\
			&				&$ (\bar{1} | 0) \; \; (\bar{3}_z | 0) \; \; (\bar{3}_z^{\, -1} | 0) \; \; (\bar{6}_z | 0) \; \; (\bar{6}_z^{\, -1} | 0) \; \; (m_z | 0) $ \\
			&				&$ (m_x | 0) \; \; (m_1 | 0) \; \; (m_{xy} | 0) \; \; (m_2 | 0) \; \; (m_y | 0) \; \; (m_3 | 0) $ \\
 27.2.90		&				&$  ( 1 | 0) \; \;  (6_z | 0) \; \; (6_z^{\, -1} | 0) \; \; (3_z | 0) \; \; (3_z^{\, -1} | 0) \; \; (2_z | 0) $ \\
 			&				&$ (2_x |  \textcolor{blue}{ \frac{1}{2}}) \; \; (2_1 |  \textcolor{blue}{ \frac{1}{2}}) \; \; (2_{xy} |  \textcolor{blue}{ \frac{1}{2}}) , \;  (2_2 |  \textcolor{blue}{ \frac{1}{2}}) \; \; (2_y |  \textcolor{blue}{ \frac{1}{2}}) \; \; (2_3 |  \textcolor{blue}{ \frac{1}{2}}) $ \\
			&				&$ (\bar{1} | 0) \; \; (\bar{3}_z | 0) \; \; (\bar{3}_z^{\, -1} | 0) \; \; (\bar{6}_z | 0) \; \; (\bar{6}_z^{\, -1} | 0) \; \; (m_z | 0) $ \\
			&				&$ (m_x |  \textcolor{blue}{ \frac{1}{2}}) \; \; (m_1 |  \textcolor{blue}{ \frac{1}{2}}) \; \; (m_{xy} |  \textcolor{blue}{ \frac{1}{2}}) \; \; (m_2 |  \textcolor{blue}{ \frac{1}{2}}) \; \; (m_y |  \textcolor{blue}{ \frac{1}{2}}) \; \; (m_3 |  \textcolor{blue}{ \frac{1}{2}}) $ \\
 27.3.91	&				&$ ( 1 | 0) \; \;   (6_z |  \textcolor{blue}{ \frac{1}{2}}) \; \; (6_z^{\, -1} |  \textcolor{blue}{ \frac{1}{2}}) \; \; (3_z | 0) \; \; (3_z^{\, -1} | 0) \; \; (2_z |  \textcolor{blue}{ \frac{1}{2}}) $\\
 			&				&$ (2_x |0) \; \; (2_1 |  \textcolor{blue}{ \frac{1}{2}}) \; \; (2_{xy} |0) , \;  (2_2 |  \textcolor{blue}{ \frac{1}{2}}) \; \; (2_y | 0) \; \; (2_3 | \textcolor{blue}{ \frac{1}{2}}) $ \\
			&				&$ (\bar{1} | 0) \; \; (\bar{3}_z | 0) \; \; (\bar{3}_z^{\, -1} |  0) \; \; (\bar{6}_z | \textcolor{blue}{ \frac{1}{2}}) \; \; (\bar{6}_z^{\, -1} | \textcolor{blue}{ \frac{1}{2}}) \; \; (m_z |  \textcolor{blue}{ \frac{1}{2}}) $ \\
			&				&$ (m_x |0) \; \; (m_1 |  \textcolor{blue}{ \frac{1}{2}}) \; \; (m_{xy} |0) \; \; (m_2 |  \textcolor{blue}{ \frac{1}{2}}) \; \; (m_y | 0) \; \; (m_3 |  \textcolor{blue}{ \frac{1}{2}}) $ \\
 27.4.92	&				&$  ( 1 | 0) \; \;  (6_z | 0) \; \; (6_z^{\, -1} | 0) \; \; (3_z | 0) \; \; (3_z^{\, -1} | 0) \; \; (2_z | 0) $ \\
 			&				&$ (2_x | 0) \; \; (2_1 | 0) \; \; (2_{xy} | 0) \; \; (2_2 | 0) \; \; (2_y | 0) \; \; (2_3 | 0) $ \\
			&				&$ (\bar{1} | \textcolor{blue}{ \frac{1}{2}}) \; \; (\bar{3}_z |  \textcolor{blue}{ \frac{1}{2}}) \; \; (\bar{3}_z^{\, -1} |  \textcolor{blue}{ \frac{1}{2}}) \; \; (\bar{6}_z |  \textcolor{blue}{ \frac{1}{2}}) \; \; (\bar{6}_z^{\, -1} |  \textcolor{blue}{ \frac{1}{2}}) \; \; (m_z |  \textcolor{blue}{ \frac{1}{2}}) $ \\
			&				&$ (m_x |  \textcolor{blue}{ \frac{1}{2}}) \; \; (m_1 |  \textcolor{blue}{ \frac{1}{2}}) \; \; (m_{xy} |  \textcolor{blue}{ \frac{1}{2}}) \; \; (m_2 |  \textcolor{blue}{ \frac{1}{2}}) \; \; (m_y |  \textcolor{blue}{ \frac{1}{2}}) \; \; (m_3 |  \textcolor{blue}{ \frac{1}{2}}) $ \\
 27.5.93	&				&$ ( 1 | 0) \; \;   (6_z | 0) \; \; (6_z^{\, -1} | 0) \; \; (3_z | 0) \; \; (3_z^{\, -1} | 0) \; \; (2_z | 0) $ \\
 			&				&$ (2_x |  \textcolor{blue}{ \frac{1}{2}}) \; \; (2_1 |  \textcolor{blue}{ \frac{1}{2}}) \; \; (2_{xy} |  \textcolor{blue}{ \frac{1}{2}}) , \;  (2_2 |  \textcolor{blue}{ \frac{1}{2}}) \; \; (2_y |  \textcolor{blue}{ \frac{1}{2}}) \; \; (2_3 |  \textcolor{blue}{ \frac{1}{2}}) $ \\
			&				&$ (\bar{1} |  \textcolor{blue}{ \frac{1}{2}}) \; \; (\bar{3}_z |  \textcolor{blue}{ \frac{1}{2}}) \; \; (\bar{3}_z^{\, -1} |  \textcolor{blue}{ \frac{1}{2}}) \; \; (\bar{6}_z |  \textcolor{blue}{ \frac{1}{2}}) \; \; (\bar{6}_z^{\, -1} |  \textcolor{blue}{ \frac{1}{2}}) \; \; (m_z |  \textcolor{blue}{ \frac{1}{2}}) $ \\
			&				&$ (m_x | 0) \; \; (m_1 | 0) \; \; (m_{xy} | 0) \; \; (m_2 | 0) \; \; (m_y | 0) \; \; (m_3 | 0) $ \\
 27.6.94	&				&$  ( 1 | 0) \; \;  (6_z |  \textcolor{blue}{ \frac{1}{2}}) \; \; (6_z^{\, -1} |  \textcolor{blue}{ \frac{1}{2}}) \; \; (3_z | 0) \; \; (3_z^{\, -1} | 0) \; \; (2_z |  \textcolor{blue}{ \frac{1}{2}}) $ \\
 			&				&$ (2_x |  \textcolor{blue}{ \frac{1}{2}}) \; \; (2_1 |0) \; \; (2_{xy} |  \textcolor{blue}{ \frac{1}{2}}) \; \; (2_2 | 0) \; \; (2_y |  \textcolor{blue}{ \frac{1}{2}}) \; \; (2_3 | 0) $ \\
			&				&$ (\bar{1} |  \textcolor{blue}{ \frac{1}{2}}) \; \; (\bar{3}_z | \textcolor{blue}{ \frac{1}{2}}) \; \; (\bar{3}_z^{\, -1} | \textcolor{blue}{ \frac{1}{2}}) \; \; (\bar{6}_z |  0) , \; ( \bar{6}_z^{\, -1} |  0) \; \; (m_z | 0) $ \\
			&				&$ (m_x | 0) \; \; (m_1 |  \textcolor{blue}{ \frac{1}{2}}) \; \; (m_{xy} | 0) \; \; (m_2 |  \textcolor{blue}{ \frac{1}{2}}) \; \; (m_y |0) \; \; (m_3 |  \textcolor{blue}{ \frac{1}{2}}) $ \\ 
 28.1.95	&23				&$  ( 1 | 0) \; \;  (3_{xyz} | 0) \; \; (3_{xy-z} | 0) \; \; (3_{-xyz} | 0) \; \; (3_{x-yz} | 0) $ \\
 			&				&$ (3_{xyz}^{-1} | 0) \; \; (3_{xy-z}^{-1} | 0) \; \; (3_{-xyz}^{-1} | 0) \; \; (3_{x-yz}^{-1} | 0) \; \; (2_z | 0) \; \; (2_y | 0) \; \; (2_x | 0) $ \\
 28.2.96	&				&$  ( 1 | 0) \; \;  (3_{xyz} |  \textcolor{blue}{ \frac{1}{3}}) \; \; (3_{xy-z} |  \textcolor{blue}{ \frac{1}{3}}) \; \; (3_{-xyz} |  \textcolor{blue}{ \frac{1}{3}}) \; \; (3_{x-yz} |  \textcolor{blue}{ \frac{1}{3}}) $ \\
 			&				&$ (3_{xyz}^{-1} |  \textcolor{blue}{ \frac{2}{3}}) (3_{xy-z}^{-1} |  \textcolor{blue}{ \frac{2}{3}}) \; \; (3_{-xyz}^{-1} |  \textcolor{blue}{ \frac{2}{3}}) \; \; (3_{x-yz}^{-1} |  \textcolor{blue}{ \frac{2}{3}}) \; \; (2_z | 0) \; \; (2_y | 0) \; \; (2_x | 0) $ \\
 29.1.97	&m$\bar{3}$		&$ ( 1 | 0) \; \;   (3_{xyz} | 0) \; \; (3_{xy-z} | 0) \; \; (3_{-xyz} | 0) \; \; (3_{x-yz} | 0) $ \\
 			&				&$ (3_{xyz}^{-1} | 0) \; \; (3_{xy-z}^{-1} | 0) \; \; (3_{-xyz}^{-1} | 0) \; \; (3_{x-yz}^{-1} | 0) \; \; (2_z | 0) \; \; (2_y | 0) \; \; (2_x | 0) $ \\
			&				&$ (\bar{1} | 0) \; \; (\bar{3}_{xyz} | 0) \; \; (\bar{3}_{xy-z} | 0) \; \; (\bar{3}_{-xyz} | 0) \; \; (\bar{3}_{x-yz} | 0) $ \\
			&				&$ (\bar{3}_{xyz}^{-1} | 0) \; \; (\bar{3}_{xy-z}^{-1} | 0) \; \; (\bar{3}_{-xyz}^{-1} | 0) \; \; (\bar{3}_{x-yz}^{-1} | 0) \; \; (m_x | 0) \; \; (m_y | 0) $ \\
			&				&$ (m_z | 0) $ \\
 29.2.98	&				&$ ( 1 | 0) \; \;   (3_{xyz} |  \textcolor{blue}{ \frac{1}{3}}) \; \; (3_{xy-z} |  \textcolor{blue}{ \frac{1}{3}}) \; \; (3_{-xyz} |  \textcolor{blue}{ \frac{1}{3}}) \; \; (3_{x-yz} |  \textcolor{blue}{ \frac{1}{3}}) $ \\
 			&				&$ (3_{xyz}^{-1} |  \textcolor{blue}{ \frac{2}{3}}) \; \; (3_{xy-z}^{-1} |  \textcolor{blue}{ \frac{2}{3}}) \; \; (3_{-xyz}^{-1} |  \textcolor{blue}{ \frac{2}{3}}) \; \; (3_{x-yz}^{-1} |  \textcolor{blue}{ \frac{2}{3}}) \; \; (2_z | 0) \; \; (2_y | 0) \; \; (2_x | 0) $ \\
			&				&$ (\bar{1} | 0) \; \; (\bar{3}_{xyz} |  \textcolor{blue}{ \frac{1}{3}}) \; \; (\bar{3}_{xy-z} |  \textcolor{blue}{ \frac{1}{3}}) \; \; (\bar{3}_{-xyz} |  \textcolor{blue}{ \frac{1}{3}}) \; \; (\bar{3}_{x-yz} |  \textcolor{blue}{ \frac{1}{3}}) \; \; (\bar{3}_{xyz}^{-1} |  \textcolor{blue}{ \frac{2}{3}}) $ \\
			&				&$ (\bar{3}_{xy-z}^{-1} |  \textcolor{blue}{ \frac{2}{3}}) \; \; (\bar{3}_{-xyz}^{-1} |  \textcolor{blue}{ \frac{2}{3}}) \; \; (\bar{3}_{x-yz}^{-1} |  \textcolor{blue}{ \frac{2}{3}}) \; \; (m_x | 0) \; \; (m_y | 0) $ \\
			&				&$ (m_z | 0) $ \\
 29.3.99	&				&$  ( 1 | 0) \; \;  (3_{xyz} | 0) \; \; (3_{xy-z} | 0) \; \; (3_{-xyz} | 0) , \;   (3_{x-yz} | 0) $ \\
 			&				&$ (3_{xyz}^{-1} | 0) \; \; (3_{xy-z}^{-1} | 0) \; \; (3_{-xyz}^{-1} | 0) \; \; (3_{x-yz}^{-1} | 0) \; \; (2_z | 0) \; \; (2_y | 0) \; \; (2_x | 0) $ \\
			&				&$ (\bar{1} |  \textcolor{blue}{ \frac{1}{2}}) \; \; (\bar{3}_{xyz} |  \textcolor{blue}{ \frac{1}{2}}) \; \; (\bar{3}_{xy-z} |  \textcolor{blue}{ \frac{1}{2}}) \; \; (\bar{3}_{-xyz} |  \textcolor{blue}{ \frac{1}{2}}) \; \; (\bar{3}_{x-yz} |  \textcolor{blue}{ \frac{1}{2}}) $ \\
			&				&$ (\bar{3}_{xyz}^{-1} |  \textcolor{blue}{ \frac{1}{2}}) \; \; (\bar{3}_{xy-z}^{-1} |  \textcolor{blue}{ \frac{1}{2}}) \; \; (\bar{3}_{-xyz}^{-1} |  \textcolor{blue}{ \frac{1}{2}}) \; \; (\bar{3}_{x-yz}^{-1} |  \textcolor{blue}{ \frac{1}{2}}) \; \; (m_x |  \textcolor{blue}{ \frac{1}{2}}) \; \; (m_y |  \textcolor{blue}{ \frac{1}{2}}) $ \\
			&				&$ (m_z |  \textcolor{blue}{ \frac{1}{2}}) $ \\
29.4.100	&				&$  ( 1 | 0) \; \;  (3_{xyz} |  \textcolor{blue}{ \frac{1}{3}}) \; \; (3_{xy-z} |  \textcolor{blue}{ \frac{1}{3}}) \; \; (3_{-xyz} |  \textcolor{blue}{ \frac{1}{3}}) \; \; (3_{x-yz} |  \textcolor{blue}{ \frac{1}{3}}) $ \\
 			&				&$ (3_{xyz}^{-1} |  \textcolor{blue}{ \frac{2}{3}}) \; \; (3_{xy-z}^{-1} |  \textcolor{blue}{ \frac{2}{3}}) \; \; (3_{-xyz}^{-1} |  \textcolor{blue}{ \frac{2}{3}}) \; \; (3_{x-yz}^{-1} |  \textcolor{blue}{ \frac{2}{3}}) \; \; (2_z | 0) \; \; (2_y | 0) \; \; (2_x | 0) $ \\
			&				&$ (\bar{1} |  \textcolor{blue}{ \frac{1}{2}}) \; \; (\bar{3}_{xyz} |  \textcolor{blue}{ \frac{5}{6}}) \; \; (\bar{3}_{xy-z} |  \textcolor{blue}{ \frac{5}{6}}) \; \; (\bar{3}_{-xyz} |  \textcolor{blue}{ \frac{5}{6}}) \; \; (\bar{3}_{x-yz} |  \textcolor{blue}{ \frac{5}{6}}) $ \\
			&				&$ (\bar{3}_{xyz}^{-1} |  \textcolor{blue}{ \frac{1}{6}}) \; \; (\bar{3}_{xy-z}^{-1} |  \textcolor{blue}{ \frac{1}{6}}) \; \; (\bar{3}_{-xyz}^{-1} |  \textcolor{blue}{ \frac{1}{6}}) \; \; (\bar{3}_{x-yz}^{-1} |  \textcolor{blue}{ \frac{1}{6}}) \; \; (m_x |  \textcolor{blue}{ \frac{1}{2}}) \; \; (m_y |  \textcolor{blue}{ \frac{1}{2}}) $ \\
			&				&$ (m_z |  \textcolor{blue}{ \frac{1}{2}}) $ \\
 30.1.101	&432				&$  ( 1 | 0) \; \;  (3_{xyz} | 0) \; \; (3_{xy-z} | 0) \; \; (3_{-xyz} | 0) \; \; (3_{x-yz} | 0) $ \\
 			&				&$ (3_{xyz}^{-1} | 0) \; \; (3_{xy-z}^{-1} | 0) \; \; (3_{-xyz}^{-1} | 0) \; \; (3_{x-yz}^{-1} | 0) \; \; (2_z | 0) \; \; (2_y | 0) \; \; (2_x | 0) $ \\
			&				&$ (4_z | 0) \; \; (4_z^{\, -1} | 0) \; \; (4_y | 0) \; \; (4_y^{\; -1} | 0) \; \; (4_x | 0) \; \; (4_x^{\; -1} | 0) $ \\
			&				&$ (2_{-xy} | 0) \; \; (2_{xy} | 0) \; \; (2_{-yz} | 0) \; \; (2_{yz} | 0) \; \; (2_{xz} | 0) \; \; (2_{-xz} | 0) $ \\
30.2.102		&				&$  ( 1 | 0) \; \;  (3_{xyz} | 0) \; \; (3_{xy-z} | 0) \; \; (3_{-xyz} | 0) \; \; (3_{x-yz} | 0) $ \\
			&				&$ (3_{xyz}^{-1} | 0) \; \; (3_{xy-z}^{-1} | 0) \; \; (3_{-xyz}^{-1} | 0) \; \; (3_{x-yz}^{-1} | 0) \; \; (2_z | 0) \; \; (2_y | 0) \; \; (2_x | 0) $ \\
			&				&$ (4_z |  \textcolor{blue}{ \frac{1}{2}}) \; \; (4_z^{\, -1} |  \textcolor{blue}{ \frac{1}{2}}) \; \; (4_y |  \textcolor{blue}{ \frac{1}{2}}) \; \; (4_y^{\; -1} |  \textcolor{blue}{ \frac{1}{2}}) \; \; (4_x |  \textcolor{blue}{ \frac{1}{2}}) \; \; (4_x^{\; -1} |  \textcolor{blue}{ \frac{1}{2}}) $ \\
			&				&$ (2_{-xy} |  \textcolor{blue}{ \frac{1}{2}}) \; \; (2_{xy} |  \textcolor{blue}{ \frac{1}{2}}) \; \; (2_{-yz} |  \textcolor{blue}{ \frac{1}{2}}) \; \; (2_{yz} |  \textcolor{blue}{ \frac{1}{2}}) \; \; (2_{xz} |  \textcolor{blue}{ \frac{1}{2}}) \; \; (2_{-xz} |  \textcolor{blue}{ \frac{1}{2}}) $ \\ 
31.1.103	&$\bar{4}$3m		&$ ( 1 | 0) \; \;   (3_{xyz} | 0) \; \; (3_{xy-z} | 0) \; \; (3_{-xyz} | 0) \; \; (3_{x-yz} | 0) $ \\
			&				&$ (3_{xyz}^{-1} | 0) \; \; (3_{xy-z}^{-1} | 0) \; \; (3_{-xyz}^{-1} | 0) \; \; (3_{x-yz}^{-1} | 0) \; \; (2_z | 0) \; \; (2_y | 0) \; \; (2_x | 0) $ \\
			&				&$ (\bar{4_z} | 0) \; \; (\bar{4}_z^{\, -1} | 0) \; \; (\bar{4}_y | 0) \; \; (\bar{4}_y^{-1} | 0) \; \; (\bar{4}_x | 0) \; \; (\bar{4}_x^{-1} | 0) $ \\
			&				&$ (m_{-xy} | 0) \; \; (m_{xy} | 0) \; \; (m_{-yz} | 0) \; \; (m_{yz} | 0) \; \; (m_{xz} | 0) \; \; (m_{-xz} | 0) $ \\
31.2.104		&				&$  ( 1 | 0) \; \;  (3_{xyz} | 0) \; \; (3_{xy-z} | 0) \; \; (3_{-xyz} | 0) \; \; (3_{x-yz} | 0) $ \\
			&				&$ (3_{xyz}^{-1} | 0) \; \; (3_{xy-z}^{-1} | 0) \; \; (3_{-xyz}^{-1} | 0) \; \; (3_{x-yz}^{-1} | 0) \; \; (2_z | 0) \; \; (2_y | 0) \; \; (2_x | 0) $ \\
			&				&$ (\bar{4_z} |  \textcolor{blue}{ \frac{1}{2}}) \; \; (\bar{4}_z^{\, -1} |  \textcolor{blue}{ \frac{1}{2}}) \; \; (\bar{4}_y |  \textcolor{blue}{ \frac{1}{2}}) \; \; (\bar{4}_y^{-1} |  \textcolor{blue}{ \frac{1}{2}}) \; \; (\bar{4}_x |  \textcolor{blue}{ \frac{1}{2}}) \; \; (\bar{4}_x^{-1} |  \textcolor{blue}{ \frac{1}{2}}) $ \\
			&				&$ (m_{-xy} |  \textcolor{blue}{ \frac{1}{2}}) \; \; (m_{xy} |  \textcolor{blue}{ \frac{1}{2}}) \; \; (m_{-yz} |  \textcolor{blue}{ \frac{1}{2}}) \; \; (m_{yz} |  \textcolor{blue}{ \frac{1}{2}}) \; \; (m_{xz} |  \textcolor{blue}{ \frac{1}{2}}) \; \; (m_{-xz} |  \textcolor{blue}{ \frac{1}{2}}) $ \\ 
32.1.105		&m$\bar{3}$m		&$ ( 1 | 0) \; \;   (3_{xyz} | 0) \; \; (3_{xy-z} | 0) \; \; (3_{-xyz} | 0) \; \; (3_{x-yz} | 0) $ \\
			&				&$ (3_{xyz}^{-1} | 0) \; \; (3_{xy-z}^{-1} | 0) \; \; (3_{-xyz}^{-1} | 0) \; \; (3_{x-yz}^{-1} | 0) \; \; (2_z | 0) \; \; (2_y | 0) \; \; (2_x | 0) $ \\
			&				&$ (4_z | 0) \; \; (4_z^{\, -1} | 0) \; \; (4_y | 0) , \; \; (4_y^{\; -1} | 0) \; \; (4_x | 0) \; \; (4_x^{\; -1} | 0) $ \\
			&				&$ (2_{-xy} | 0) \; \; (2_{xy} | 0) \; \; (2_{-yz} | 0) \; \; (2_{yz} | 0) \; \; (2_{xz} | 0) \; \; (2_{-xz} | 0) $ \\
			&				&$ (\bar{1} | 0) \; \; (\bar{4_z} | 0) \; \; (\bar{4}_z^{\, -1} | 0) \; \; (\bar{4}_y | 0) \; \; (\bar{4}_y^{-1} | 0) \; \; (\bar{4}_x | 0) $ \\
			&				&$ (\bar{4}_x^{-1} | 0) \; \; (\bar{3}_{xyz} | 0) \; \; (\bar{3}_{xy-z} | 0) \; \; (\bar{3}_{-xyz} | 0) \; \; (\bar{3}_{x-yz} | 0) \; \; (\bar{3}_{xyz}^{-1} | 0) $ \\
			&				&$ (\bar{3}_{xy-z}^{-1} | 0) \; \; (\bar{3}_{-xyz}^{-1} | 0) \; \; (\bar{3}_{x-yz}^{-1} | 0) \; \; (m_z | 0) \; \; (m_y | 0) \; \; (m_x | 0) $ \\
			&				&$ (m_{-xy} | 0) \; \; (m_{xy} | 0) \; \; (m_{-yz} | 0) \; \; (m_{yz} | 0) \; \; (m_{xz} | 0) \; \; (m_{-xz} | 0) $ \\
32.2.106	&				&$ ( 1 | 0) \; \;   (3_{xyz} | 0) \; \; (3_{xy-z} | 0) \; \; (3_{-xyz} | 0) \; \; (3_{x-yz} | 0) $ \\
			&				&$ (3_{xyz}^{-1} | 0) \; \; (3_{xy-z}^{-1} | 0) \; \; (3_{-xyz}^{-1} | 0) \; \; (3_{x-yz}^{-1} | 0) \; \; (2_z | 0) \; \; (2_y | 0) \; \; (2_x | 0) $ \\
			&				&$ (4_z |  \textcolor{blue}{ \frac{1}{2}}) \; \; (4_z^{\, -1} |  \textcolor{blue}{ \frac{1}{2}}) \; \; (4_y |  \textcolor{blue}{ \frac{1}{2}}) \; \; (4_y^{\; -1} |  \textcolor{blue}{ \frac{1}{2}}) \; \; (4_x |  \textcolor{blue}{ \frac{1}{2}}) \; \; (4_x^{\; -1} |  \textcolor{blue}{ \frac{1}{2}}) $ \\
			&				&$ (2_{-xy} |  \textcolor{blue}{ \frac{1}{2}}) \; \; (2_{xy} |  \textcolor{blue}{ \frac{1}{2}}) \; \; (2_{-yz} |  \textcolor{blue}{ \frac{1}{2}}) \; \; (2_{yz} |  \textcolor{blue}{ \frac{1}{2}}) \; \; (2_{xz} |  \textcolor{blue}{ \frac{1}{2}}) \; \; (2_{-xz} |  \textcolor{blue}{ \frac{1}{2}}) $ \\
			&				&$ (\bar{1} | 0) \; \; (\bar{4_z} |  \textcolor{blue}{ \frac{1}{2}}) \; \; (\bar{4}_z^{\, -1} |  \textcolor{blue}{ \frac{1}{2}}) , \;  (\bar{4}_y |  \textcolor{blue}{ \frac{1}{2}}) \; \; (\bar{4}_y^{-1} | \textcolor{blue}{ \frac{1}{2}}) \; \; (\bar{4}_x |  \textcolor{blue}{ \frac{1}{2}}) $ \\
			&				&$ (\bar{4}_x^{-1} |  \textcolor{blue}{ \frac{1}{2}}) \; \; (\bar{3}_{xyz} | 0) \; \; (\bar{3}_{xy-z} | 0) \; \; (\bar{3}_{-xyz} | 0) \; \; (\bar{3}_{x-yz} | 0) \; \; (\bar{3}_{xyz}^{-1} | 0) $ \\
			&				&$ (\bar{3}_{xy-z}^{-1} | 0) \; \; (\bar{3}_{-xyz}^{-1} | 0) \; \; (\bar{3}_{x-yz}^{-1} | 0) \; \; (m_z | 0) \; \; (m_y | 0) \; \; (m_x | 0) $ \\
			&				&$ (m_{-xy} |  \textcolor{blue}{ \frac{1}{2}}) \; \; (m_{xy} |  \textcolor{blue}{ \frac{1}{2}}) \; \; (m_{-yz} |  \textcolor{blue}{ \frac{1}{2}}) \; \; (m_{yz} |  \textcolor{blue}{ \frac{1}{2}}) \; \; (m_{xz} |  \textcolor{blue}{ \frac{1}{2}}) \; \; (m_{-xz} |  \textcolor{blue}{ \frac{1}{2}}) $ \\
32.3.107	&				&$  ( 1 | 0) \; \;  (3_{xyz} | 0) \; \; (3_{xy-z} | 0) \; \; (3_{-xyz} | 0) \; \; (3_{x-yz} | 0) $ \\
			&				&$ (3_{xyz}^{-1} | 0) \; \; (3_{xy-z}^{-1} | 0) \; \; (3_{-xyz}^{-1} | 0) \; \; (3_{x-yz}^{-1} | 0) \; \; (2_z | 0) \; \; (2_y | 0) $ \\
			&				&$ (2_x | 0) \; \; (4_z | 0) \; \; (4_z^{\, -1} | 0) \; \; (4_y | 0) \; \; (4_y^{\; -1} | 0) \; \; (4_x | 0) $ \\
			&				&$ (4_x^{\; -1} | 0) \; \; (2_{-xy} | 0) \; \; (2_{xy} | 0) \; \; (2_{-yz} | 0) \; \; (2_{yz} | 0) \; \; (2_{xz} | 0) $ \\
			&				&$ (2_{-xz} | 0) \; \; (\bar{1} |  \textcolor{blue}{ \frac{1}{2}}) \; \; (\bar{4_z} |  \textcolor{blue}{ \frac{1}{2}}) \; \; (\bar{4}_z^{\, -1} |  \textcolor{blue}{ \frac{1}{2}}) \; \; (\bar{4}_y |  \textcolor{blue}{ \frac{1}{2}}) \; \; (\bar{4}_y^{-1} |  \textcolor{blue}{ \frac{1}{2}}) $ \\
			&				&$ (\bar{4}_x |  \textcolor{blue}{ \frac{1}{2}}) \; \; (\bar{4}_x^{-1} |  \textcolor{blue}{ \frac{1}{2}}) \; \; (\bar{3}_{xyz} |  \textcolor{blue}{ \frac{1}{2}}) \; \; (\bar{3}_{xy-z} |  \textcolor{blue}{ \frac{1}{2}}) $ \\
			&				&$ (\bar{3}_{-xyz} |  \textcolor{blue}{ \frac{1}{2}}) \; \; (\bar{3}_{x-yz} |  \textcolor{blue}{ \frac{1}{2}}) \; \; (\bar{3}_{xyz}^{-1} |  \textcolor{blue}{ \frac{1}{2}}) \; \; (\bar{3}_{xy-z}^{-1} |  \textcolor{blue}{ \frac{1}{2}}) \; \; (\bar{3}_{-xyz}^{-1} |  \textcolor{blue}{ \frac{1}{2}}) $ \\
			&				&$ (\bar{3}_{x-yz}^{-1} |  \textcolor{blue}{ \frac{1}{2}}) \; \; (m_z |  \textcolor{blue}{ \frac{1}{2}}) \; \; (m_y |  \textcolor{blue}{ \frac{1}{2}}) $ \\
			&				&$ (m_x |  \textcolor{blue}{ \frac{1}{2}}) \; \; (m_{-xy} |  \textcolor{blue}{ \frac{1}{2}}) \; \; (m_{xy} |  \textcolor{blue}{ \frac{1}{2}}) \; \; (m_{-yz} |  \textcolor{blue}{ \frac{1}{2}}) \; \; (m_{yz} |  \textcolor{blue}{ \frac{1}{2}}) \; \; (m_{xz} |  \textcolor{blue}{ \frac{1}{2}}) $ \\
			&				&$ (m_{-xz} |  \textcolor{blue}{ \frac{1}{2}}) $ \\
32.4.108	&				&$  ( 1 | 0) \; \;  (3_{xyz} | 0) \; \; (3_{xy-z} | 0) \; \; (3_{-xyz} | 0) \; \; (3_{x-yz} | 0) $ \\
			&				&$ (3_{xyz}^{-1} | 0) \; \; (3_{xy-z}^{-1} | 0) \; \; (3_{-xyz}^{-1} | 0) \; \; (3_{x-yz}^{-1} | 0) \; \; (2_z | 0) \; \; (2_y | 0) $ \\
			&				&$ (2_x | 0) \; \; (4_z |  \textcolor{blue}{ \frac{1}{2}}) \; \; (4_z^{\, -1} |  \textcolor{blue}{ \frac{1}{2}}) \; \; (4_y |  \textcolor{blue}{ \frac{1}{2}}) \; \; (4_y^{\; -1} |  \textcolor{blue}{ \frac{1}{2}}) \; \; (4_x |  \textcolor{blue}{ \frac{1}{2}}) $ \\
			&				&$ (4_x^{\; -1} |  \textcolor{blue}{ \frac{1}{2}}) \; \; (2_{-xy} |  \textcolor{blue}{ \frac{1}{2}}) \; \; (2_{xy} |  \textcolor{blue}{ \frac{1}{2}}) \; \; (2_{-yz} |  \textcolor{blue}{ \frac{1}{2}}) \; \; (2_{yz} |  \textcolor{blue}{ \frac{1}{2}}) \; \; (2_{xz} |  \textcolor{blue}{ \frac{1}{2}}) $ \\
			&				&$ (2_{-xz} |  \textcolor{blue}{ \frac{1}{2}}) \; \; (\bar{1} |  \textcolor{blue}{ \frac{1}{2}}) \; \; (\bar{4_z} | 0) \; \; (\bar{4}_z^{\, -1} | 0) \; \; (\bar{4}_y | 0) \; \; (\bar{4}_y^{-1} | 0) $ \\
			&				&$ (\bar{4}_x | 0) \; \; (\bar{4}_x^{-1} | 0) \; \; (\bar{3}_{xyz} |  \textcolor{blue}{ \frac{1}{2}}) \; \; (\bar{3}_{xy-z} |  \textcolor{blue}{ \frac{1}{2}}) \; \; (\bar{3}_{-xyz} |  \textcolor{blue}{ \frac{1}{2}}) \; \; (\bar{3}_{x-yz} |  \textcolor{blue}{ \frac{1}{2}}) $ \\
			&				&$ (\bar{3}_{xyz}^{-1} |  \textcolor{blue}{ \frac{1}{2}}) \; \; (\bar{3}_{xy-z}^{-1} |  \textcolor{blue}{ \frac{1}{2}}) \; \; (\bar{3}_{-xyz}^{-1} |  \textcolor{blue}{ \frac{1}{2}}) \; \; (\bar{3}_{x-yz}^{-1} |  \textcolor{blue}{ \frac{1}{2}}) \; \; (m_z |  \textcolor{blue}{ \frac{1}{2}}) \; \; (m_y |  \textcolor{blue}{ \frac{1}{2}}) $ \\
			&				&$ (m_x |  \textcolor{blue}{ \frac{1}{2}}) \; \; (m_{-xy} | 0) \; \; (m_{xy} | 0) \; \; (m_{-yz} | 0) \; \; (m_{yz} | 0) \; \; (m_{xz} | 0) $ \\
			&				&$ (m_{-xz} | 0) $ \\

\end{longtable}

\pagebreak

\section*{Table 2 - Formulas to generate spatio-temporal point groups}

In this table, formulas used to generate both crystallographic as well as non-crystallographic spatio-temporal point groups are shown.

The first and second columns specify in Sch{\"o}nflies and International \cite{Aroyo2016} notation respectively, the point group from which the corresponding spatio-temporal group is derived. The third column shows a coset representative of each spatio-temporal group. In these groups, $n$ denotes an n-fold rotation, or a rotation by angle $\phi= \frac{2\pi}{n}$ in radians. In the case of limiting (or infinite) point groups, $\infty$ is used to denote an $\infty$-fold rotation, while $\phi$ denotes the corresponding infinitesimal angle of rotation in radians.

A shorthand notation is used to list the coset representative of each group. For example, the set of elements generated by an n-fold rotation with zero time translation, that is $\{(n^0|0),(n^1|0),(n^2|0),...,(n^{n-1}|0) \}$, is represented by $(n^j|0)|_{j=0,1,...,n-1}$. In the case of limiting point groups, the index $j$ is dropped. For example the set of elements generated by an $\infty$-fold rotation with zero time translation is represented simply by $(\infty|0)$ ... .\\

\begin{longtable}{llll}
Sch{\"o}nflies Notation	& International Notation    & Spatio-temporal Group   \\
\hline
$C_n$	&n			&$ (n^{j} | 0)  |_{ j = 0, 1, ... , n-1} $ \\
		&			&$ (n^{j} | \textcolor{blue}{ \frac{j}{n}}) |_{ j = 0, 1, ... , n-1} $ \\
		&			&. \\
		&			&. \\
		&			&. \\
		&			&$ (n^{j} | \; \textcolor{blue}{ \frac{j(n-1)}{n}}) |_{ j = 0, 1, ... , n-1} $ \\
$C_{nv}$	&nmm (even n)	&$ (n^{j} | 0) |_{ j = 0, 1, ... , n-1} \; \; (m_x | 0 ) \; \; ...  \; \; (m_{xy} | 0 ) \; \; ... \; \; $ \\
		&		&$ (n^{j} | 0) |_{ j = 0, 1, ... , n-1} \; \; (m_x | \textcolor{blue}{ \frac{1}{2}}) \; \; ...  \; \; (m_{xy} | \textcolor{blue}{ \frac{1}{2}}) \; \; ... \; \; $ \\
		&		&$ (n^{j} | 0) |_{ j = 0, 2, ... , n-2} \; \; (n^{j} | \textcolor{blue}{ \frac{1}{2}}) |_{ j = 1, 3, ... , n-1} \; \; (m_x | 0 ) \; \; ... \; \; $ \\ 
		&		&$ (m_{xy} | \textcolor{blue}{ \frac{1}{2}}) \; \; ... \; \; $ \\
		&		&$ (n^{j} | 0) |_{ j = 0, 2, ... , n-2} \; \; (n^{j} | \textcolor{blue}{ \frac{1}{2}}) |_{ j = 1, 3, ... , n-1} \; \; (m_x | \textcolor{blue}{ \frac{1}{2}}) \; \; ...  \; \; $ \\ 
		&		&$(m_{xy} | 0 ) \; \; ... \; \; $ \\
		&nm (odd n)	&$ (n^{j} | 0) |_{ j = 0, 1, ... , (n-1)} \; \; (m_x | 0 ) \; \; ... $ \\
		&		&$ (n^{j} | 0) |_{ j = 0, 1, ... , (n-1)} \; \; (m_x | \textcolor{blue}{ \frac{1}{2}}) \; \; ... \; \; $ \\
$C_{nh}$	&n/m (even n) &$ (n^{j} | 0) |_{ j = 0, 1, ... , n-1} \; \; (n^{j} m_z | 0) |_{j = 0, 1, ... , n} $ \\
		&		&$ (n^{j} | \textcolor{blue}{ \frac{j}{n}}) |_{ j = 0, 1, ... , n-1} \; \; (n^{j} m_z | \textcolor{blue}{ \frac{j}{n}})  |_{j = 0, 1, ... , n} $ \\ 
		&			&. \\
		&			&. \\
		&			&. \\
		&		&$ (n^{j} | \textcolor{blue}{ \frac{j(n-1)}{n}}) |_{ j = 0, 1, ... , n-1} \; \; (n^{j} m_z | \textcolor{blue}{ \frac{j(n-1)}{n}}) |_{j = 0, 1, ... , n} $ \\
		&$\overline{2n}$ (odd n)	&$ (n^{j} | 0) |_{j = 0, 1, ... , n-1} \; \; (\overline{n}^{j} | 0) |_{j = 0, 1, ... , n-1} \; \; (m_x | 0) \; \; ... \; \; $ \\
		&	&$ (n^{j} | \textcolor{blue}{ \frac{j}{n}}) |_{j = 0, 1, ... , n-1} \; \; (\overline{n}^{j} | \textcolor{blue}{ \frac{j}{n}}) |_{j = 0, 1, ... , n-1} \; \; (m_x | 0) \; \; ... \; \; $ \\
		&			&. \\
		&			&. \\
		&			&. \\
		&	&$ (n^{j} | \textcolor{blue}{ \frac{j}{n}}) |_{j = 0, 1, ... , n-1} \; \; (\overline{n}^{j} | \textcolor{blue}{ \frac{j(n-1)}{n}}) |_{j = 0, 1, ... , n-1} \; \; $ \\
		&	&$ (m_x | 0) \; \; ... \; \; $ \\
$S_{2n}$	&$\overline{2n}$ (even n) &$ (2n^{j} | 0) |_{j = 0, 1, ... , n-1} \; \; (\overline{2n}^{j} | 0) |_{j = 0, 1, ... , n-1} $ \\
		&	&$ (2n^{j} | \textcolor{blue}{ \frac{j}{n}}) |_{j = 0, 1, ... , n-1} \; \; (\overline{2n}^{j} | \textcolor{blue}{ \frac{j}{n}}) |_{j = 0, 1, ... , n-1} \; \; $ \\
		&			&. \\
		&			&. \\
		&			&. \\
		& &$ (2n^{j} | \textcolor{blue}{ \frac{j(n-1)}{n}}) |_{j = 0, 1, ... , n-1} \; \; \{ (\overline{2n}^{j} | \textcolor{blue}{ \frac{j(n-1)}{n}}) |_{ j = 0, 1, ... , n-1} \; \; $ \\
		&$\overline{n}$	(odd n) &$ (\overline{2n}^{j} | 0) |_{j = 0, 1, ... , n-1} \; \; $\\
		&	 &$ (\overline{2n}^{j} | \textcolor{blue}{ \frac{j}{n}}) |_{j = 0, 1, ... , n-1} \; \; $ \\
		&			&. \\
		&			&. \\
		&			&. \\
		& &$ (\overline{2n}^{j} | \textcolor{blue}{ \frac{j(n-1)}{n}})  |_{ j = 0, 1, ... , n-1} \; \; $ \\ 
$D_n$	&n22 (even n)	&$ (n^j | 0) |_{j = 0, 1, ... , n-1} \; \; (2_x | 0) \; \; ... \; \; (2_{xy} | 0) \; \; ... \;  $ \\
		&		&$ (n^j | 0) |_{j = 0, 1, ... , n-1} \; \; (2_x | \textcolor{blue}{ \frac{1}{2}}) \; \; ... \; \; (2_{xy} | \textcolor{blue}{ \frac{1}{2}}) \; \; ... \; $ \\
		&		&$ (n^{j} | 0) |_{ j = 0, 2, ... , n-2} \; \; (n^{j} | \textcolor{blue}{ \frac{1}{2}}) |_{ j = 1, 3, ... , n-1} (2_x | 0) \; \; ... \; \; $ \\ 
		&		&$ (2_{xy} | \textcolor{blue}{ \frac{1}{2}}) \; \; ... \; $ \\
		&		&$ (n^{j} | 0) |_{ j = 0, 2, ... , n-2} \; \; (n^{j} | \textcolor{blue}{ \frac{1}{2}}) |_{ j = 1, 3, ... , n-1} (2_x | \textcolor{blue}{ \frac{1}{2}}) \; \; ... \; \; $ \\
		&		&$ (2_{xy} | 0) \; \; ... \;  $ \\
		&n2 (odd n)	&$ (n^j | 0) |_{j = 0, 1, ... , n-1} \; \; (2_x | 0) \; \; ... \; $ \\
		&		&$ (n^j | 0) |_{j = 0, 1, ... , n-1} \; \; (2_x | \textcolor{blue}{ \frac{1}{2}}) \; \; ... \; $ \\
$D_{nh}$	&n/mmm (even n) &$ (n^{j} | 0)  |_{j = 0, 1, ... , n-1} \; \; (2_x | 0) \; \; ... \; \; (2_{xy} | 0) \; \; ... \; \; (m_z | 0) \; \; $ \\
		&			&$ (n^j m_z | 0)  |_{j = 1, 2, ... , n-1} \; \; (2_x m_z | 0) \; \; ... \; \; (2_{xy} m_z | 0) \; \; ... \; $ \\
		&	&$ (n^{j} | 0) |_{j = 0, 1, ... , n-1} \; \; (2_x | \textcolor{blue}{ \frac{1}{2}}) \; \; ... \; \; (2_{xy} | \textcolor{blue}{ \frac{1}{2}}) \; \; ... \; \: (m_z | 0) \; \; $ \\
		&			&$ (n^j m_z | 0)  |_{j = 0, 1, ... , n-1} \; \; (2_x m_z | \textcolor{blue}{ \frac{1}{2}}) \; \; ... \; \; (2_{xy} m_z | \textcolor{blue}{ \frac{1}{2}}) \; \; ... \; $ \\
		&	&$ (n^{j} | 0) |_{ j = 0, 2, ... , n-2} \; \; (n^{j} | \textcolor{blue}{ \frac{1}{2}}) |_{ j = 1, 3, ... , n-1} (2_x | 0) \; \; ... \; \; $ \\
		&	&$ (2_{xy} | \textcolor{blue}{ \frac{1}{2}}) \; \; ... \; \; (m_z | 0) \; \; (n^j m_z | \textcolor{blue}{ \frac{1}{2}}) |_{j = 0, 1, ... , n-1} \; \; $ \\
		&			&$ (2_x m_z | 0) \; \; ... \; \; (2_{xy} m_z | \textcolor{blue}{ \frac{1}{2}}) \; \; ... \; $ \\
		&	&$ (n^{j} | 0) |_{ j = 0, 2, ... , n-2} \; \; (n^{j} | \textcolor{blue}{ \frac{1}{2}}) |_{ j = 1, 3, ... , n-1} (2_x | \textcolor{blue}{ \frac{1}{2}}) \; \; ... \; \; $ \\
		&			&$ (2_{xy} | 0) \; \; ... \; \; (m_z | 0) \; \; (n^j m_z | \textcolor{blue}{ \frac{1}{2}}) |_{j = 0, 1, ... n-1} \; \; $ \\
		&	&$  (2_x m_z | \textcolor{blue}{ \frac{1}{2}}) \; \; ... \; \; (2_{xy} m_z | \textcolor{blue}{ \frac{1}{2}}) \; \; ... \; $ \\
	 	&	&$ (n^{j} | 0) |_{j = 0, 1, ... n-1} \; \; (2_x | 0) \; \; ... \; \; (2_{xy} | \textcolor{blue}{ \frac{1}{2}}) \; \; ... \; \; (m_z | \textcolor{blue}{ \frac{1}{2}}) \; \; $ \\
		&			&$ (n^j m_z | \textcolor{blue}{\frac{1}{2}}) |_{j = 0, 1, ... n-1} \; \; (2_x m_z | \textcolor{blue}{ \frac{1}{2}}) \; \; ... \; \; (2_{xy} m_z | \textcolor{blue}{ \frac{1}{2}}) \; \; ... \; $ \\
		&	&$ (n^{j} | 0) |_{j = 0, 1, ... n-1} \; \; (2_x | \textcolor{blue}{ \frac{1}{2}}) \; \; ... \; \; (2_{xy} | \textcolor{blue}{ \frac{1}{2}}) \; \; ... \; \; (m_z | \textcolor{blue}{ \frac{1}{2}}) \; \; $ \\
		&			&$ (n^j m_z | \textcolor{blue}{ \frac{1}{2}}) |_{j = 0, 1, ... n-1} \; \; (2_x m_z | 0) \; \; ... \; \; (2_{xy} m_z | 0) \; \; ... \; \; $ \\
		&	&$ (n^{j} | 0) |_{ j = 0, 2, ... , n-2} \; \; (n^{j} | \textcolor{blue}{ \frac{1}{2}}) |_{ j = 1, 3, ... , n-1} (2_x | 0) \; \; ... \; \; $ \\
		&	&$ (2_{xy} | \textcolor{blue}{ \frac{1}{2}}) \; \; ... \; \; (m_z | \textcolor{blue}{ \frac{1}{2}}) \; \; (n^j m_z | \textcolor{blue}{ \frac{1}{2}}) |_{j = 0, 1, ... n-1} \; \; $ \\
		&			&$ (2_x m_z | \textcolor{blue}{ \frac{1}{2}}) \; \; ... \; \; (2_{xy} m_z | 0) \; \; ... \; $ \\
		&	&$ (n^{j} | 0) |_{ j = 0, 2, ... , n-2} \; \; (n^{j} | \textcolor{blue}{ \frac{1}{2}}) |_{ j = 1, 3, ... , n-1} (2_x | \textcolor{blue}{ \frac{1}{2}}) \; \; ... \; \; $ \\
		&			&$ (2_{xy} | 0) \; \; ... \; \; (m_z | \textcolor{blue}{ \frac{1}{2}}) \; \; (n^j m_z | 0) |_{j = 0, 1, ... n-1} \; \; $ \\
		&	&$ (2_x m_z | 0) \; \; ... \; \; (2_{xy} m_z | \textcolor{blue}{ \frac{1}{2}}) \; \; ... \; \; $ \\
		&$\overline{2n}$m2 (odd n)	&$ (n^{j} | 0)  |_{ j = 0, 1, ... , n-1} \; \; (2_x | 0) \; \; ... \; \; (n^j m_z | 0) |_{ j = 0, 1, ... , n-1} \; \; $ \\
		&			&$ (2_x m_z | 0) \; \; ... \; $ \\
		& 	&$ (n^{j} | 0) |_{j = 0, 1, ... , n-1} \; \; (2_x | \textcolor{blue}{ \frac{1}{2}}) \; \; ... \; \; (n^j m_z | 0) |_{j = 0, 1, ... n-1} \; \; $ \\
		&			&$ (2_x m_z | \textcolor{blue}{ \frac{1}{2}}) \; \; ... \; $ \\
		&	&$ (n^{j} | 0) |_{j = 0, 1, ... , n-1}  \; \; (2_x | 0) \; \; ... \; \; (n^j m_z | \textcolor{blue}{ \frac{1}{2}}) |_{j = 0, 1, ... n-1} \; \; $ \\
		&			&$(2_x m_z | \textcolor{blue}{ \frac{1}{2}}) \; \; ... \; $ \\
		&	&$(n^{j} | 0) |_{j = 0, 1, ... , n-1} \; \; (2_x | \textcolor{blue}{ \frac{1}{2}}) \; \; ... \; \; (n^j m_z | \textcolor{blue}{ \frac{1}{2}}) |_{j = 0, 1, ... n-1} \; \; $ \\
		&			&$ (2_x m_z | 0) \;\; ... \; $ \\			
$D_{nd}$	&$\overline{2n}$2m	(even n) &$ (\overline{2n}^j | 0) |_{j = 0, 1, ... 2n} \; \; (m_x | 0) \; \; ... \; \; (2_{xy} | 0) \; \; ... \;  $ \\
		&	&$ (\overline{2n}^j | 0) |_{j = 0, 1, ... 2n} \; \; (m_x | \textcolor{blue}{ \frac{1}{2}}) \; \; ... \; \; (2_{xy} | \textcolor{blue}{ \frac{1}{2}}) \; \; ... \; $ \\
		&	&$ (\overline{2n}^j | 0) |_{j = 0, 1, ... 2n} \; \; (m_x | 0) \; \; ... \; \; (2_{xy} | \textcolor{blue}{ \frac{1}{2}}) \; \; ... \; $ \\
		&	&$ (\overline{2n}^j | 0) |_{j = 0, 1, ... 2n} \; \; (m_x | \textcolor{blue}{ \frac{1}{2}}) \; \; ... \; \; (2_{xy} | 0) \; \; ... \;  $ \\
		&$\overline{n}$m	(odd n) &$ (\overline{2n}^j | 0) |_{j = 0, 1, ... 2(n-1)} \; \; (m_x | 0) \; \; ... \; \; (2_{xy} | 0) \; \; ... \;  $ \\
		&	&$ (\overline{2n}^j | 0) |_{j = 0, 1, ... 2(n-1)} \; \; (m_x | \textcolor{blue}{ \frac{1}{2}}) \; \; ... \; \; (2_{xy} | \textcolor{blue}{ \frac{1}{2}}) \; \; ... \; $ \\
		&	&$ (\overline{2n}^j | 0) |_{j = 0, 1, ... 2(n-1)} \; \; (m_x | 0) \; \; ... \; \; (2_{xy} | \textcolor{blue}{ \frac{1}{2}}) \; \; ... \; $ \\
		&	&$ (\overline{2n}^j | 0) |_{j = 0, 1, ... 2(n-1)} \; \; (m_x | \textcolor{blue}{ \frac{1}{2}}) \; \; ... \; \; (2_{xy} | 0) \; \; ... \;  $ \\
$C_{\infty}$	&$\infty$	&$ (1 | 0) \; \; (\infty | 0) \; \; ... \; $ \\	
			&		&$ (1 | 0) \; \; (\infty | \textcolor{blue}{ \frac{\phi}{2\pi}}) \; \; ... \; $ \\
			&		&$ (1 | 0) \; \; (\infty | \textcolor{blue}{ \frac{-\phi}{2\pi}}) \; \; ... \; $ \\
$C_{\infty v}$	&$\infty$mm	&$ (1 | 0) \; \; (\infty | 0) \; \; ... \; \; (\overline{\infty} | 0) \; \; ... \; \; (m_x | 0) \; \;  ... \; \; $ \\
			&			&$ (1 | 0) \; \; (\infty | 0) \; \; ... \; \; (\overline{\infty} | 0) \; \; ... \; \; (m_x | \textcolor{blue}{ \frac{1}{2}}) \; \;  ... \; \; $ \\
$C_{\infty h}$	&$\infty$/m	&$ (1 | 0) \; \; (\infty | 0) \; \; ... \; \; (\overline{\infty} | 0) \;  \; ... \; $ \\
			&		&$ (1 | 0) \; \; (\infty | \textcolor{blue}{ \frac{\phi}{2 \pi }}) \; \; ... \; \; (\overline{\infty} | \textcolor{blue}{\frac{-\phi}{2 \pi}}) \;  \; ... \; $ \\
			&		&$ (1 | 0) \; \; (\infty | \textcolor{blue}{ \frac{-\phi}{2 \pi }}) \; \; ... \; \; (\overline{\infty} | \textcolor{blue}{ \frac{\phi}{2 \pi}}) \; \; ... \;  $ \\
			&		&$ (1 | 0) \; \; (\infty | \textcolor{blue}{ \frac{\phi}{2 \pi }}) \; \; ... \; \; (\overline{\infty} | \textcolor{blue}{ \frac{\phi}{2 \pi }}) \; \; ... \; $ \\
			&		&$ (1 | 0) \; \; (\infty | \textcolor{blue}{ \frac{-\phi}{2 \pi }}) \; \; ... \; \; (\overline{\infty} | \textcolor{blue}{ \frac{-\phi}{2 \pi }}) \; \; ... \; $ \\
$D_{ \infty}$	&$\infty$2		&$ (1 | 0) \; \; (\infty | 0) \; \; ... \; \; (2_x | 0) ... $ \\
			&		&$ (1 | 0) \; \; (\infty | 0) \; \; ... \; \; (2_x | \textcolor{blue}{ \frac{1}{2}}) ... $ \\
$D_{ \infty h}$   &$\infty$/mmm	&$ (1 | 0) \; \; (\infty | 0) \; \; ... \; \; (m_x | 0) \; \; ... \; \; (\overline{1} | 0) \; \; (\overline{\infty} | 0) \; \; ... \; \; $ \\
			&				&$ (2_x | 0) \; \;  ... \; \; $ \\
			&		&$ (1 | 0) \; \; (\infty | 0) \; \; ... \; \; (m_x | \textcolor{blue}{ \frac{1}{2}}) \; \; ... \; \; (\overline{1} | 0) \; \; (\overline{\infty} | 0) \; \; ... \; \; $ \\
			&				&$ (2_x | \textcolor{blue}{ \frac{1}{2}}) \; \; ... $ \\
			&		&$  (1 | 0) \; \; (\infty | 0) \; \; ... \; \; (m_x | 0) \; \; ... \; \; (\overline{1} | \textcolor{blue}{ \frac{1}{2}}) \; \; (\overline{\infty} | \textcolor{blue}{ \frac{1}{2}}) \; \; ... \; \; $ \\
			&				&$(2_x | \textcolor{blue}{ \frac{1}{2}}) \; \; ... $ \\
			&		&$  (1 | 0) \; \; (\infty | 0) \; \; ... \; \; (m_x | \textcolor{blue}{ \frac{1}{2}}) \; \; ... \; \; (\overline{1} | \textcolor{blue}{ \frac{1}{2}}) \; \; (\overline{\infty} | \textcolor{blue}{ \frac{1}{2}}) \; \; ... \; \; $ \\
			&				&$ (2_x | 0) \; \; ... $ \\
$K$			&$\infty \infty$		&$ (1 | 0) \; \;  (\infty | 0) \; \; ... \; \; $ \\
$K_h$		&$\infty \infty$m	&$ (1 | 0) \; \;  (\infty | 0) \; \; ... \; \; (\overline{\infty} | 0) \; \; ... \; \; (\overline{1} | 0) $ \\
			&		&$ (1 | 0) \; \;  (\infty | 0) \; \; ... \; \; (\overline{\infty} | \textcolor{blue}{ \frac{1}{2}}) \; \; ... \; \; (\overline{1} | \textcolor{blue}{ \frac{1}{2}}) $ \\

\end{longtable}

\pagebreak

\section{Acknowledgements}

The authors thank Latham Boyle and Kendrick Smith for sharing their notes on platonic orbits and the character table method to obtain spatio-temporal groups. The authors also thank Jeremy Karl Cockcroft for giving permission to modify and use the stereographic projections from his online course on powder diffraction \cite{Stereo}.

\printbibliography

\end{document}